\documentclass[11pt]{article}
\pdfoutput=1 

\usepackage{jheppub} 

\usepackage[T1]{fontenc} 
\usepackage[title]{appendix}

\makeatother

\newcommand{\tr}{{\textrm{tr}}}

\begin{document}

	\vspace*{4mm}
	
	\thispagestyle{empty}

	\begin{center}
		
			\begin{minipage}{20cm}
				\begin{center}
					\hspace{-5cm }
					\LARGE
					\sc
					Anomaly induced transport
					\\
					\hspace{-5cm} from symmetry breaking in holography
				\end{center}
			\end{minipage}
			\\[15mm]
			
			\renewcommand{\thefootnote}{\fnsymbol{footnote}}

			{\large
				Ashis Tamang$^{\, a}$ \footnote{ashismaniram@gmail.com},
				Nishal Rai$^{\, b, c}$ \footnote{nishalrai10@gmail.com},
				Karl Landsteiner$^{\, c}$ \footnote{karl.landsteiner@csic.es},
				Eugenio Meg\'{\i}as$^{\, e}$ \footnote{emegias@ugr.es} \,
			}\\[12mm]
		\end{center}
		\noindent

		${}^{a}\!$
		\textit{Department of Physics, SRM University Sikkim, Upper Tadong, Sikkim, 737102,  India}

		${}^{b}\!$
		\textit{Department of Physics, St. Joseph's College, North Point, Darjeeling, 734104, India}

                ${}^{c}\!$
		\textit{Center for Astrophysics, Gravitation and Cosmology (CAGC),} \\
        \indent \, \textit{SRM University Sikkim, Upper Tadong, Sikkim, 737102,  India}

		${}^{d}\!$
		\textit{Instituto de F\'{\i}sica Te\'orica UAM/CSIC, C/ Nicol\'as Cabrera 13-15,} \\
		\indent \; \textit{Universidad Aut\'onoma de Madrid, Cantoblanco, 28049 Madrid, Spain} 
		
		${}^{e}\!$
		\textit{Departamento de F{\'\i}sica At\'omica, Molecular y Nuclear and} \\
		\indent \; \textit{Instituto Carlos I de F{\'\i}sica Te\'orica y Computacional,}  \\
		\indent \; \textit{Universidad de Granada, Avenida de Fuente Nueva s/n, E-18071 Granada, Spain}

		\vspace*{4mm}

		\begin{center}
			{  \bf  Abstract }
		\end{center}
		\begin{minipage}{15cm}
			
			We study the transport properties of relativistic fluids induced by quantum anomalies in presence of explicit symmetry breaking. To this end we consider a holographic Einstein-Maxwell model in 5 dimensions with pure gauge and a mixed gauge-gravitational Chern-Simons terms, coupled with a scalar field. To study the chiral vortical effects and the energy transport sector, apart from the chiral magnetic effects, we have considered the full backreaction of the gauge field on the metric. We have studied the anomalous effects by using Kubo formulae involving correlators of the charged currents and the energy current. Our findings reveal that, in the presence of explicit symmetry breaking, anomaly-induced transport phenomena can extend beyond anomalous currents and affect non-anomalous sectors as well. In particular, we find that all the conductivities display a distinct sensitivity to the mass parameter controlling the symmetry breaking, thus reflecting the interplay between anomaly coefficients and explicit symmetry breaking terms. These findings highlight the role played by pure gauge and mixed gauge-gravitational anomalies in holographic transport, and their importance for strongly coupled systems with broken symmetries.

			\vspace{0.5cm}
		\end{minipage}

		\tableofcontents

		\section{Introduction}
		
		In recent times, the relationship between conformal field theories (CFT) and gravity theories on specific backgrounds has been extensively explored. The most prominent and thoroughly studied example of this correspondence is the  one between ${\mathcal N} = 4$ supersymmetric Yang-Mills (SYM) theory in the large $N$ limit and type IIB string theory in the AdS$_5$ $\times$ S$^5$ space \cite{Maldacena:1997re,Witten:1998qj}. When the field theory is at strong coupling, the string theory simplifies to classical supergravity, making it possible to compute all correlation functions in the field theory.

		Anomaly induced transport phenomena are extremely active areas of research that arise in the context of relativistic field theories of chiral fermions, going beyond the scope of perturbation theory \cite{Kharzeev:2015znc,Bastianelli:2006rx}. These anomalies have been instrumental in shaping the modern understanding of relativistic hydrodynamics \cite{Landsteiner:2012kd}. Quantum anomalies, such as the chiral anomaly, give rise to novel transport phenomena such as the chiral magnetic (CME) and the chiral vortical (CVE) effect that arise at finite temperature and  chemical potential. The CME is observed in systems with chiral imbalance, and it is responsible for the generation of an electric current induced by an applied magnetic field and parallel to it~\cite{Fukushima:2008xe}, while the CVE is the generation of an electric current induced by a vortex in the medium and parallel to the vorticity vector. These are examples of anomaly induced transport phenomena, meaning that quantum corrections to classical symmetries are the ultimate responsible of them, and they represent fascinating phenomena arising in certain quantum systems. Moreover, an important property of these phenomena is that the anomaly induced currents are non-dissipative, i.e. they don't contribute to entropy production~\cite{Son:2009tf}.

		The anomaly-induced effects have practical implications in high-energy physics and in condensed matter systems. In high energy physics the search for signatures of anomaly induced transport in heavy ion collisions is an on-going endeavor and has culminated in the iso-bar program at RHIC and the LHC \cite{STAR:2009wot,Kharzeev:2019zgg}. These collisions could lead to event-by-event parity violation, which is a significant departure from the expected symmetry properties of the system. The CVE can also be important in heavy ion collisions especially due to the proven presence of large vorticity \cite{STAR:2017ckg}. In condensed matter physics the CME is at the origin of the observed large enhancement of the longitudinal conductivity of Weyl and Dirac semi-metals subject to a magnetic field~\cite{Li:2014bha,Gooth:2017mbd}. Similar anomalous chiral transport effects have been studied also in chiral superfluids with vorticity~\cite{Bevan:1997pqv,Nissinen:2021gke}.
		
		In recent years, these anomalous effects have been explored using holographic methods, providing significant insights~\cite{Bhattacharyya:2007vjd,Erdmenger:2008rm,Landsteiner:2011iq,Rai:2023nxe,Gallegos:2024qxo}. One such study is \cite{Megias:2016ery}, where a holographic model with a pure Chern-Simons term was considered. This work examined the role of the axial anomaly in the hydrodynamics of relativistic fluids in the presence of an external electric field. In this study, the anomalous conductivities were computed using Kubo formulae. A particularly interesting result from this research is the identification of a novel phenomenon where anomalies can induce transport effects not only in anomalous currents but also in non-anomalous ones, i.e., those with a vanishing divergence, when symmetries are explicitly broken. This phenomenon was studied at strong coupling in a holographic Einstein-Maxwell model in five dimensions, supplemented with a pure gauge Chern-Simons term. Symmetry breaking was introduced through a scalar field dual to an operator ${\cal O}$ with $\dim {\cal O} = 3$. Our goal in the present work is to extend the study for other anomalous coefficients, like the chiral vortical conductivity. In addition, we will study the role played by the mixed gauge-gravitational anomaly in the generation of the non-anomalous currents. In order to achieve these goals, one should consider the following ingredients: {\it i)} make the computation including the full backreaction of the gauge fields onto the metric tensor, and {\it ii)} add a mixed gauge-gravitational Chern-Simons terms in the action of the model. More specifically we will consider a model with three symmetries, two of them can be identified with the usual vector and axial $U(1)$ symmetries of Dirac fermions. Of these only the axial symmetry is anomalous. In addition we consider a non-anomalous symmetry\footnote{To be precise we note that also the vector symmetry is non-anomalous. The vector gauge field does however appear in the Chern-Simons term and this means that it has a {\em covariant} but not a {\em consistent} anomaly. In contrast the gauge field corresponding to the additional symmetry does not appear in the Chern-Simons terms. For brevity we will refer to it as non-anomalous.}, and a symmetry breaking scalar field that is simultaneously charged under the axial and the non-anomalous symmetry. The  boundary value $M$ of the scalar field is the symmetry breaking parameter.
		
		The structure of this manuscript is organized as follows. We summarize in section~\ref{sec:constitutive} the main results discussed in the literature about the constitutive relations related to anomalous transport, and motivate the introduction of a non-anomalous symmetry. In section~\ref{sec:action} we present the holographic model under consideration, the ansatz used for the solutions of the different fields, and derive the fully backreacted numerical solution for the background. The model comprises of different background parameters such as $\mu_v$, $\mu_a$ and $\mu_w$, which correspond to chemical potentials for the vector, axial and non-anomalous symmetry respectively. In addition to this, there are two more background parameters: the black hole temperature $T$, and $M$ which is the symmetry breaking parameter. We fix the position of the horizon in the numerical calculation. Subsequently, in section~\ref{sec:flut}, we consider the perturbations on both the metric and the gauge fields on top of the background computed in section~\ref{sec:back_sol}. Next, in section~\ref{sec:kubo_formulae} we present the Kubo formulae associated with these conductivities. In section~\ref{sec: result} we present our results: first, in subsection~\ref{subsec:numerical_method} we start by furnishing the method used to solve numerically the equations of motion of the fluctuations, followed by subsection~\ref{sec: result1} where we present the comparison between our numerics and the analytic results at $M=0$ to confirm the reliability of the numerical method used. In subsection~\ref{sec: result2}, we will present the results of the $\mu_v$-dependence of the conductivities that induce the non-anomalous current $\vec{J}_w$, for different values of $M/(\pi T)$. Furthermore, we present in subsection~\ref{sec: result3} the results of the conductivities vs. $M/(\pi T)$ in two different cases: {\it i)} fixed non-vanishing chemical potentials, and {\it ii)}~vanishing chemical potentials, i.e. $\left( \mu_v = \mu_a = \mu_w = 0 \right)$. We summarize our findings and provide an outlook for future directions in section~\ref{sec:discussion}. Finally, appendix~\ref{sec:appendix_EoM} contains the equations of motion for the background and for the fluctuations, appendix~\ref{sec:App_finite_M} collects the numerical results for all the conductivities as a function of the chemical potentials, and appendix~\ref{sec:App_large_M} presents the analytical results for the conductivities in the large $M$ limit.

		\section{Constitutive relations and (non)anomalous transport}
		\label{sec:constitutive}
		
		We will summarize in this section the transport effects that will be studied in the rest of the manuscript. Let us consider a theory of $N$ chiral fermions transforming under a global symmetry group $G$ generated by matrices $(T_s)^f\,{}_g$. The chemical potential for the fermion $\Psi^f$ is given by $\mu^f= \sum_s q_s^f \mu_s$, while the Cartan generator is $H_s = q^f_s \, \delta^f\,{}_g$ where $q^f_s$ are the charges. The general form of the anomalous induced currents by a magnetic field is
		\begin{equation}
			\vec{J}_a  = \sigma^B_{ab}  \, \vec{B}_b \,, \label{eq:CME}
		\end{equation}
		where $\vec{B}_b$ is the magnetic field corresponding to symmetry $b$. The 1-loop computation of the chiral magnetic conductivity by using the Kubo formalism leads to~\cite{Kharzeev:2009pj,Landsteiner:2011cp,Chowdhury:2015pba}
		\begin{equation}
			\sigma^B_{ab} = d_{abc} \frac{\mu^c}{4\pi^2}  \,, \qquad  d_{abc} = \frac 1 2 [ \tr( T_a \{ T_b, T_c\} )_R -  \tr( T_a \{ T_b, T_c\} )_L ] \,,
		\end{equation}
		where $d_{abc}$ is the group theoretic factor related to the axial anomaly, which typically appears in the computation of the anomalous triangle diagram corresponding to three non-abelian gauge fields coupled to a chiral fermion. The subscripts $R$, $L$ stand for the contributions of right-handed and left-handed fermions. Anomalies are responsible for a non-vanishing value of the divergence of the current, that reads in this case~\cite{Kimura:1969iwz}
		\begin{equation}
			D_\mu J_a^\mu= \epsilon^{\mu\nu\rho\lambda} \frac{d_{abc}}{32\pi^2} F^b_{\mu\nu} F^c_{\rho\lambda}    \,.
		\end{equation}
		Let us particularize eq.~(\ref{eq:CME}) to the symmetry group $U_V(1) \times U_A(1)$. Then there are vector $(v)$ and axial $(a)$ currents induced by the magnetic field of the vector field, i.e.
		\begin{equation}
			\vec{J}_v = \frac{\mu_a}{2\pi^2} \vec{B}_v  \,, \qquad \vec{J}_a = \frac{\mu_v}{2\pi^2} \vec{B}_v  \,, \label{eq:CMECSE}
		\end{equation}
		which correspond to the CME and the chiral separation effect (CSE), respectively. At this point, the following question arises: is it possible to get a {\it chiral magnetic effect} for a non-anomalous symmetry~$w$? This means to have an induced current in symmetry $w$, i.e.
		\begin{equation}
			\vec{J}_w \ne 0 \qquad \textrm{with} \qquad    d_{wab} = d_{awb} = d_{abw} = 0 \qquad \forall a,b \,.
		\end{equation}
		In the rest of the manuscript we will study the possibility that anomalies can induce transport also in non-anomalous currents through magnetic fields and the vorticity vector.

		\section{Holographic action}
		\label{sec:action}
		
		In this work, we consider a holographic model that includes both a pure gauge and mixed gauge-gravitational Chern-Simons terms in the action, which mimics the axial anomaly and gravitational contribution to the axial anomaly respectively. The action reads \cite{Landsteiner:2013aba,Megias:2016ery}
		\begin{eqnarray}
			\mathcal S &=& \dfrac{1}{16 \pi G} \int d^5x\sqrt{-g}
			\Big[
			R + 2\Lambda -\dfrac{1}{4}  F_V^{2} \nonumber -\dfrac{1}{4}  F_A^{2} \nonumber -\dfrac{1}{4}  F_W^{2} \nonumber \nonumber \\&&+ \epsilon^{MNPQR} A_M \left(\dfrac{\kappa}{3}(F_A)_{NP}(F_A)_{QR}+{\kappa}(F_V)_{NP}(F_V)_{QR}+\lambda R^A\,{}_{BNP}R^B\,{}_{AQR} \right)\Big] \nonumber \\&&+ \mathcal S_{\rm GH} + \mathcal S_{\rm CSK}   \,,  
			\label{act}
		\end{eqnarray}
		where $\Lambda = 6 / \ell^2$  is the cosmological constant with $\ell$ the radius of AdS. The stress tensors are defined as $(F_S)_{MN}=\partial_M S_N-\partial_N S_M$ for $S = V, A, W$, with $V$ and $A$ denoting the anomalous vector and axial gauge fields, and $W$ representing an extra non-anomalous gauge field. The remaining parameters are given by $\kappa$ and $\lambda$, which denotes the Chern-Simons couplings associated with the anomalous sectors. The Levi-Civita symbol is defined by $\epsilon(rtxyz) = +1$ whereas the Levi-Civita tensor is defined by $\epsilon^{MNPQR}=\epsilon(MNPQR)/\sqrt{-g}$. The boundary terms are given by
		\begin{eqnarray}
			\mathcal S_{\rm GH} &=& \dfrac{1}{8 \pi G}\int_\partial d^4x    \sqrt{h} K \,,\\
			\mathcal S_{\rm CSK} &=& -\dfrac{1}{2 \pi G}\int_\partial d^4x    \sqrt{h} K\lambda n_M\epsilon^{MNPQR}A_NK_{PL}D_QK^L_R \,,  
		\end{eqnarray}
		and they correspond, respectively, to the Gibbons-Hawking term required to have a well-defined variational problem, and a term induced by the mixed gauge-gravitational anomaly. The values of the couplings $\kappa$ and $\lambda$ are given in terms of the anomalous coefficients of the field theory as~\cite{Landsteiner:2011iq}
		\begin{equation}
			\label{eq:kappa-lambda_value}
			\frac{1}{16\pi G}\kappa = -\frac{1}{16\pi^2} \,,  \qquad \frac{1}{16\pi G}
			\lambda = -\frac{1}{384\pi^2} \,.
		\end{equation}
		
		The inclusion of both Chern-Simons terms makes this holographic framework sufficiently rich to study anomaly-induced transport. Since only the axial gauge potential $A$ appears in the Chern-Simons term only the axial symmetry is anomalous. The metric $g$ and the vector gauge field $V$
		appear however with their corresponding field strength and Riemann tensor. In contrast the gauge field $W$ does not appear at all in the Chern-Simons term. In the subsequent subsections, we will extend our model further by introducing explicit symmetry breaking terms, enabling us to investigate how anomalies affect non-anomalous sectors rather than just anomalous sectors of the theory.

		\subsection{Holographic model with symmetry breaking}
		\label{sec:sym_break}

		A key objective of this work is to examine the possibility that the constitutive relation for $\langle J_w^\mu \rangle$ receives anomalous contributions despite the symmetry being non-anomalous. To explore this possibility, we extend our model of eq.~(\ref{act}) by introducing explicit symmetry breaking via a scalar field, $S_\phi$ \cite{Megias:2016ery}, so that 
		\begin{equation}
			\mathcal S_{\textrm{tot}} = \mathcal S + \mathcal S_\phi  \,,
		\end{equation}
		where, $\mathcal S$ is the bulk action comprising the usual Maxwell kinetic terms and both pure and mixed Chern-Simons contributions, while the additional scalar contribution is
		\begin{equation}
			\mathcal S_\phi = \frac{1}{16\pi G} \int d^5x \sqrt{-g} \bigg(  - |D_M \phi|^2 - m^2\phi^2  \bigg) \,, \qquad  D_M \phi = \left[ \partial_M -i(A_M - W_M) \right] \phi  \,. \label{eq:Sphi}
		\end{equation} 
		Here $\phi$ is a scalar field with a tachyonic bulk mass $m^2 = \Delta (\Delta -4) / \ell^2$, and $0 \le \Delta \le 4$. $\mathcal S_\phi$ produces an explicit breaking of $a$ and $w$ symmetries via the scalar field~$\phi$. From the AdS/CFT dictionary, the model is the holographic dual of a Conformal Field Theory (CFT) with a deformation
		\begin{equation}
			{\cal L} = {\cal L}^{\rm CFT} + M \, {\cal O} \,,  
		\end{equation}
		where ${\cal O}$ is an operator dual of the scalar field with $\dim {\cal O} = \Delta$, and $M$ is the source of the operator with $\dim M = 4-\Delta$. The near boundary expansion of~$\phi$ reads
		\begin{equation}
			\phi(r) = \phi_{4-\Delta} \left( \frac{r}{\ell} \right) ^{\Delta-4} + \phi_{\Delta} \left( \frac{r}{\ell} \right)^{-\Delta} + \cdots \,, \qquad r \to \infty \,, \label{eq:phir}
		\end{equation}
		where $\phi_{4-\Delta}$ is interpreted as the source $M$, and $\phi_\Delta$ as the condensate $\langle {\cal O}\rangle$. In the following we will choose $\Delta = 3$, so that the bulk mass is $m^2 = -3/\ell^2$.

		\subsection{Numerical solution for the Background }
		\label{sec:back_sol}
		In order to account for the chiral vortical effects within our model, we will consider the full backreaction of the gauge field onto the metric tensor, and include a mixed gauge-gravitational Chern-Simons term in the action. The equation of motion for the background fields, obtained from 
		\begin{equation}
			\frac{\delta \mathcal S}{\delta \varphi} = 0\,, \qquad 
			\varphi \in \{ g_{MN}, V^M, A^M, W^M, \phi \} \,,
		\end{equation}
		for the action (\ref{act}) together with (\ref{eq:Sphi}), can be solved by considering the following ansatz for the metric
		\begin{equation}
			ds^2 = - \frac{r^2}{\ell^2}  f(r)dt^2 + e^{\chi(r)} \frac{\ell^2}{r^2} \frac{dr^2}{f(r)} + \frac{r^2}{\ell^2} \left( dx^2 + dy^2 + dz^2\right)   \,,
		\end{equation}
		and the background gauge fields
		\begin{equation}
			V = V_t(r)dt \,, \qquad A = A_t(r)dt \,, \qquad W = W_t(r)dt  \,. 
		\end{equation}
		The chemical potentials are computed as $\mu_s \equiv S_t(r\to\infty) - S_t(r_h)$ with $S = V, A, W$. With these changes, the fields have the following near boundary expansion
		\begin{equation}
			\lim_{r\to\infty} r \cdot \phi(r) = M \,, \quad \lim_{r\to\infty} V_t(r) = \mu_v  \,, \quad \lim_{r\to\infty} A_t(r) = \mu_a  \,, \quad  \lim_{r\to\infty} W_t(r) = \mu_w  \,, \label{eq:nearboundary}
		\end{equation}
		where we have assumed in these expressions that the gauge fields must vanish on the horizon: $S_t(r_h) = 0$ $\; (S= V, A, W)$. It is convenient to define the following new fields $A_\pm \equiv A \pm W$, so that the covariant derivative writes~$D_M\phi = \left[ \partial_M -i(A_-)_M \right] \phi$. Consequently, the chemical potentials take the form $\mu_\pm \equiv \mu_a  \pm \mu_w$.
		
		The equations of motion of the background for the fields $V$, $A_+$, $A_-$, $\phi$ and $\chi$ are shown in appendix~\ref{subsec:appendix_background}, cf. eqs.~(\ref{eq:Vt})-(\ref{eq:chi}). These constitute a system of six coupled differentials equations that can be solved numerically with  appropriate boundary conditions given by (\ref{eq:nearboundary}) together with regularity near the horizon. First we will consider the change of variable $u=\frac{r_h^2}{r^2}$, where $r_h$ is the position of the horizon. In the $u$ variable the horizon is at $u_h=1$. In the massless case $(M=0)$, the gauge fields and the blackening factor turn out to be
		\begin{eqnarray}
			S_t(u) &=&  \mu_s \left( 1 - u \right)  \,,\qquad\qquad  (S = V, A, W) \,, \\
			f(u) &=& 1 - \frac{\ell^2 \mathcal M}{r_h^4} u^2 + \frac{\ell^2 \sum_s Q_s^2}{r_h^6} u^3 \,,
		\end{eqnarray}
		where
		\begin{equation}
			\mathcal M = \frac{r_h^4}{\ell^2} + \frac{\sum_s Q_s^2}{r_h^2} \,, \qquad Q_s = \frac{\mu_s r_h^2}{\sqrt{3}}   \,,\qquad\qquad  (s = v, a, w) \,,   \label{eq:Mass_Charges}
		\end{equation}
		are the mass and charges of the black hole, respectively. Then, the Hawking temperature is given by
		\begin{equation}
			T = \frac{r_h}{2 \pi \ell^2} \left| f^\prime(u) \right|_{u=1} = \frac{r_h}{\pi \ell^2} \left( 1 - \frac{\ell^2}{6 r_h^2} \left( \mu_v^2 + \mu_a^2 + \mu_w^2 \right) \right) \,. 
		\end{equation}
		The dependence of $T$ on $\mu_w$ is a consequence of the fact that the non-anomalous field $W$ affects the background even in the absence of symmetry breaking, as it is obvious from the action of the model, eq.~(\ref{act}). However, this does not necessarily imply the existence of transport for the current~$\vec{J}_w$.

\begin{figure}[htbp]
			\centering
			\includegraphics[width=.46\textwidth]{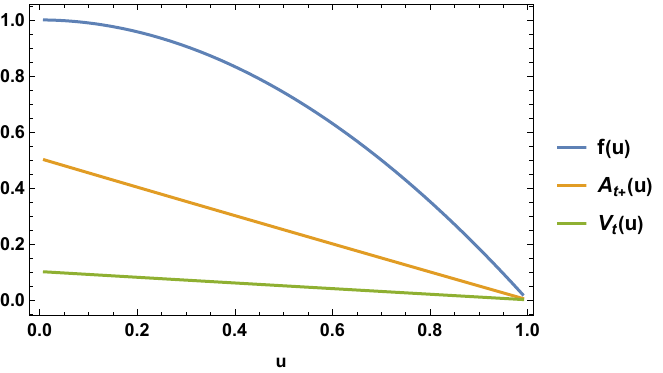}
			\qquad
			\includegraphics[width=.46\textwidth]{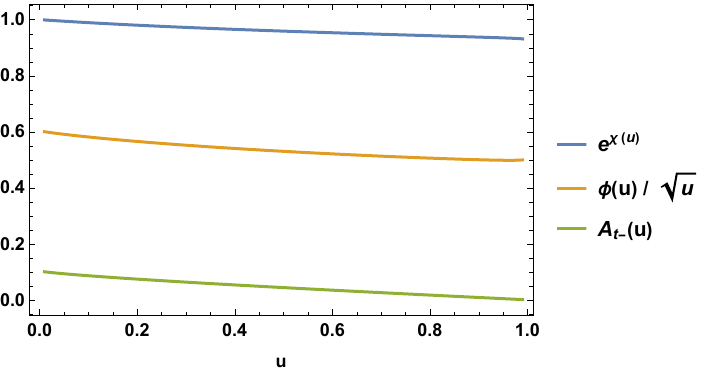}
			\caption{Plot showing the dependence of the background with $u$. Color code: Blue -  $f(u)$, Orange - $A_{t+} (u)$,  and Green - $V_t(u)$ (left panel); Blue - $e^{\chi(u)}$, Orange - $\phi(u) / \sqrt{u}$, and Green - $A_{t-} (u)$ (right panel). We have set $\mu_v = 0.1$, $\mu_a=0.3$, $\mu_w=0.2$ and $M=0.6$. \label{fig:i}}
		\end{figure}

		In the massive case $(M > 0)$, we must resort to numerical methods to solve the coupled differential equations: we have used the shooting method, shooting from the horizon $u_h=1$ towards the boundary $u=0$. We plot in fig.~\ref{fig:i} the numerical solutions of all the background fields as a function of $u$. In the rest of the manuscript we will set $\ell = 1$ unless otherwise stated.
		
		\section{Fluctuations}
		\label{sec:flut}
		We will now consider perturbations of fields, where the background is set by the numerical solution as shown in fig.~\ref{fig:i}. To study the effect of anomalies, we split the metric and gauge fields into background and a linear perturbation part, i.e.   
		\begin{align}
			g_{MN} ={g^{(0)}}_{MN} + \epsilon h_{MN}\,,\,\,\, \qquad   \mathcal S_\mu = S_\mu + \epsilon s_\mu \,, 
		\end{align}
		where $S_\mu$ refers to the background solutions computed in section~\ref{sec:back_sol} for any of the fields $V$, $A_\pm$ (or equivalently $V$, $A$ and $W$), and $s_\mu$ are the corresponding fluctuations $v, a_\pm \equiv a \pm w$ (or equivalently $v$, $a$ and $w$). 
		Then, we will follow the general procedure of Fourier mode decomposition~\cite{Amado:2011zx}
		\begin{eqnarray}
			h_{MN}(u, x^\nu) &=& \int \dfrac{d^4p}{(2\pi)^4} \, h_{MN}(u) \, e^{-i\omega t+i \vec{p} \cdot \vec{x}} \,,\\
			s_{\mu}(u, x^\nu) &=& \int \dfrac{d^4p}{(2\pi)^4} \, s_{\mu}(u) \, e^{-i\omega t+i \vec{p} \cdot \vec{x}}\,.
		\end{eqnarray}
		Without the loss of generality, we restrict our analysis to perturbations characterized by frequency $\omega$ and momentum $\vec{p}$ directed along the $z$-axis.  Since we are interested in computing correlators at zero frequency, we can set to zero the frequency-dependent parts in the equations and solve the system up to first order in $p \equiv |\vec{p}|$. To compute the retarded correlators and hence the conductivities, we need to solve the equations of motion for the fluctuations of the metric and gauge fields. The explicit expressions of these equations is provided in appendix~\ref{subsec:appendix_fluctuations}.

		\section{Constitutive relations and Kubo formulae}
		\label{sec:kubo_formulae}
		
		The transport coefficients for the chiral vortical and magnetic effects, $\sigma_V$ and $\sigma_B$, have been studied in a wide variety of methods which includes Kubo formulae \cite{Amado:2011zx,Landsteiner:2012kd,Chowdhury:2015pba}, fluid gravity correspondence \cite{Bhattacharyya:2007vjd,Erdmenger:2008rm,Banerjee:2008th,Megias:2013joa}, diagrammatic methods \cite{Manes:2012hf} and the  partition function formalism \cite{Banerjee:2012iz,Megias:2014mba,Jensen:2012jy,Jensen:2012jh}. In this work we use Kubo formalism to study the chiral vortical effect of anomalous conductivities. We will study the possibility that, under certain circumstances, this effect might be present in non-anomalous currents as well.
		
		In the following we will use the notation $J^i_S \in \{ J^i_v, \, J^i_a, \, J^i_w , J^i_\varepsilon \}$, where we include the vector and axial currents, the current related to the $w$ symmetry, and the energy current $\left( J^i_\varepsilon \equiv T^{0i} \right)$, respectively. The constitutive relations for the charged and energy currents write
		\begin{eqnarray}
			\vec{J}_v &=& \sigma^B_{vv} \, \vec{B}_v + \sigma^B_{va} \, \vec{B}_a + \sigma^B_{vw} \, \vec{B}_w + \sigma^V_{v} \, \vec{\Omega} \,,  \label{eq:Jv_const}  \\
			\vec{J}_a &=& \sigma^B_{av} \, \vec{B}_v + \sigma^B_{aa} \, \vec{B}_a + \sigma^B_{aw} \, \vec{B}_w + \sigma^V_{a} \, \vec{\Omega} \,,  \label{eq:Ja_const}  \\
			\vec{J}_w &=& \sigma^B_{wv} \, \vec{B}_v + \sigma^B_{wa} \, \vec{B}_a + \sigma^B_{ww} \, \vec{B}_w + \sigma^V_{w} \, \vec{\Omega} \,, \label{Jw} \\
			\vec{J}_\varepsilon &=& \sigma^B_{\varepsilon v} \, \vec{B}_v + \sigma^B_{\varepsilon a} \, \vec{B}_a + \sigma^B_{\varepsilon w} \, \vec{B}_w + \sigma^V_{\varepsilon} \, \vec{\Omega} \,.  \label{eq:Je_const}
		\end{eqnarray}
		We have denoted by $\sigma^B_{sb}$ the chiral magnetic conductivity which induces the  current $\vec{J}_s$  as a response of the magnetic field $\vec{B}_b$ associated to symmetry $b \in \{ v, a, w\}$. By the same way, $\sigma^V_{s}$ denotes the chiral vortical conductivity which induces the current $\vec{J}_s$ as a response of the vorticity vector $\vec{\Omega}$. It is understood that $s \in \{ v, a, w, \varepsilon \}$ in these expressions, so that these correspond either to conductivities for the charged currents ($\sigma^B_{sb}$ and $\sigma^V_s$  with $s \in \{v, a, w\}$) or for the energy current ($\sigma^B_{\varepsilon b}$ and $\sigma^V_\varepsilon$).
		
		The Kubo formula for the chiral magnetic conductivity for the dual field theory of the holographic model involves the retarded correlators of two currents, as it was derived in~\cite{Kharzeev:2009pj}. By using the compact notation introduced above, the Kubo formulae for the chiral magnetic conductivities write
		\begin{equation}
			\sigma^B_{s b} = \lim_{p_k \rightarrow 0}\dfrac{i}{2 p_k}\sum_{i,j} \epsilon_{ijk}\langle J_s^{i} J_b^{j}\rangle\vert_{\omega =0} \,, \qquad   \left(s \in \{ v, a, w, \varepsilon \}  \,, \;  b \in \{ v, a, w \} \right) \,.
			\label{eqku2}
		\end{equation}
		The Kubo formula for the chiral vortical conductivity involves the energy-momentum tensor~\cite{Amado:2011zx}. In the present case, the Kubo formulae write
		\begin{equation}
			\sigma^V_s =\lim_{p_k\rightarrow 0}\dfrac{i}{2 p_k}\sum_{i,j} \epsilon_{ijk}\langle J_s^i J_\varepsilon^j \rangle\vert_{\omega =0}\,, \qquad \left( s \in \{ v, a, w, \varepsilon \} \right) \,. \label{eqku1}
		\end{equation}
		After taking the hermitian conjugate of the correlators, and using the hermiticity of the currents, one has that $\langle J_s^i(p) J_{s^\prime}^j(-p) \rangle = \langle J_{s^\prime}^j(p) J_s^j(-p) \rangle$ in the zero frequency limit. Then it can be deduced the following relations between the chiral magnetic and chiral vortical conductivities
		\begin{equation}
			\sigma^B_{\varepsilon b} = \sigma^V_{b} \,, \qquad (b \in \{v, a, w \}) \,.  \label{eq:sigmaV_symmetric}
		\end{equation}
		Similar relations between the chiral magnetic conductivities can be deduced by the same argument, and they write
		\begin{equation}
			\sigma^B_{ab} = \sigma^B_{ba}  \,, \qquad (a, b \in \{ v, a, w \} ) \,. \label{eq:sigmaB_symmetric}
		\end{equation}
		In an obvious unified notation for the chiral magnetic and chiral vortical conductivities, $\sigma_{s s^\prime}$, the properties of eqs.~(\ref{eq:sigmaV_symmetric}) and (\ref{eq:sigmaB_symmetric}) imply that this matrix is symmetric.

		\section{Results for the conductivities}
		\label{sec: result}
		
		Before presenting the numerical results for the conductivities, it is worth providing the numerical method we opted to solve the system of equations for the fluctuations.
		
		\subsection{Numerical method}
		\label{subsec:numerical_method}
		
		To solve the coupled differential equations for the fluctuations we have used the pseudo-spectral method~\cite{trefethen2000spectral, boyd2000spectral}. To get the solutions we expand the fluctuations as a sum of Chebyshev polynomials in $u$ direction given by 
		\begin{equation}
			(a_\pm)_i=\sum_{m=0}^{N-1} (a_\pm^{m})_i T_m (2u-1) \,,  \quad
			v_i=\sum_{m=0}^{N-1}v_i^m T_m (2u-1) \,,  \quad
			h_{it}=\sum_{m=0}^{N-1}h_{it}^m T_m (2u-1) \,,
			\label{cheb}
		\end{equation}
		where $(a_\pm^m)_i$, $v_i^m$ and $h_{it}^m$ are the coefficients of the polynomials in (\ref{cheb}) with ($i=x,y$). These polynomials are inserted into the equation of motion for the fluctuations, and the collocation points for $u=0$ to $u=1$ are chosen in Gauss-Lobatto grid. By choosing the appropriate number of grid points corresponding to the number of coefficients one can solve the set of algebraic equations in terms of these coefficients to get the solution for the fluctuations.
		
		In the following subsections we will present our results. Note that we are working at zero frequency and in the limit of zero momentum: $\omega=0$ and $p\rightarrow 0$.
		
		\subsection{Results of the conductivities for $M=0$}
		\label{sec: result1}
		
\begin{figure}[htbp]
			\centering
			\includegraphics[width=.4\textwidth]{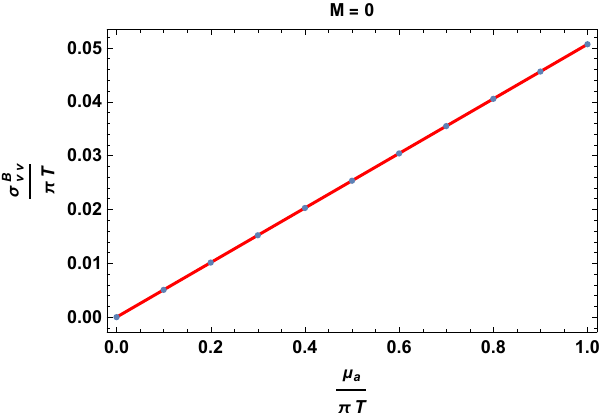}
			\qquad
			\includegraphics[width=.4\textwidth]{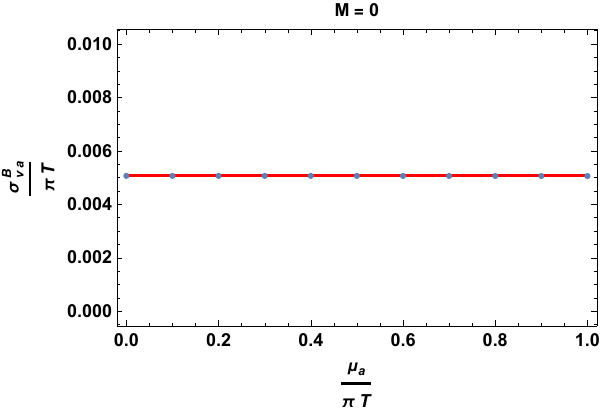}
			\caption{Plots of the conductivities $\sigma^{B}_{v v} = \sigma^{B}_{a a}$ and $\sigma^{B}_{v a} =\sigma^{B}_{a v}$ in the massless case $(M = 0)$ as a function of $\mu_a/(\pi T)$. Solid red lines correspond to the analytic results of ~(\ref{eq:sigmaAB_ana})-(\ref{eq:sigmaAV_ana}), while blue dots stand for the numerical results computed by using the method of sec.~\ref{subsec:numerical_method}.  We have set $\mu_v/(\pi T)=0.1$. \label{fig:coryvM0}}
		\end{figure}

\begin{figure}[htbp]
			\centering
			\includegraphics[width=.4\textwidth]{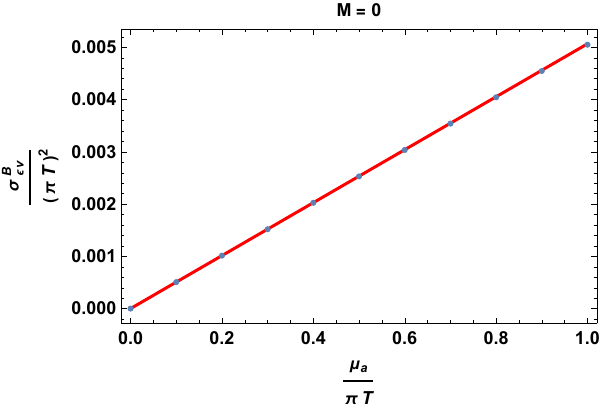}
			\includegraphics[width=.4\textwidth]{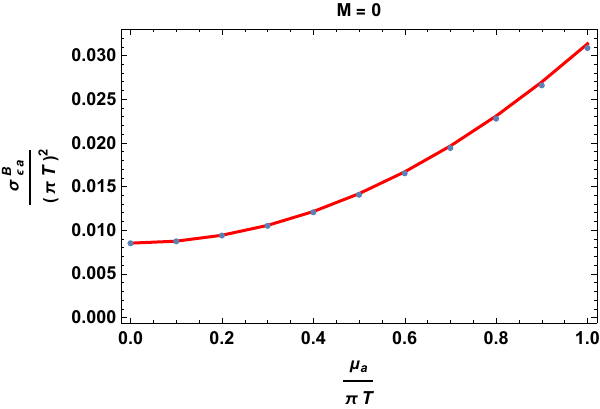}
			\caption{Plots of the conductivities $\sigma^{B}_{\varepsilon v} = \sigma^{V}_{v}$ and $\sigma^{B}_{\varepsilon a} = \sigma^{V}_{a}$ in the massless case $(M = 0)$  as a function of $\mu_a/(\pi T)$. See fig.~\ref{fig:coryvM0} for further details. \label{fig:coryaM0}}
		\end{figure}

\begin{figure}[htbp]
			\centering
			\includegraphics[width=.43\textwidth]{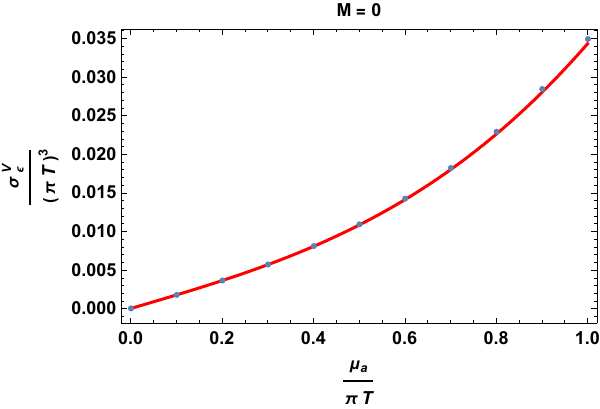}
			\caption{Plot of the conductivity $\sigma^{V}_{\varepsilon}$   in the massless case $(M = 0)$  as a function of $\mu_a/(\pi T)$. See fig.~\ref{fig:coryvM0} for further details. \label{fig:corytM0}}
		\end{figure}

		The case $M=0$ has been studied in the literature. The analytical results in this case are (see Ref.~\cite{Landsteiner:2013aba} and references therein)
		\begin{eqnarray}
			\sigma^{B}_{s v}(0) &=& \frac{1}{2\pi^2} \left\{
			\begin{array}{ll}
				\mu_a \,, & \qquad (s = v) \\
				\mu_v  \,, &  \qquad (s = a) \\
				\mu_v \mu_a \,, & \qquad (s = \varepsilon)
			\end{array}
			\right.,   \label{eq:sigmaAB_ana} \\
			\sigma^{B}_{s a}(0) &=& \frac{1}{2\pi^2} \left\{
			\begin{array}{ll}
				\mu_v \,, & \qquad (s = v) \\
				\mu_a \,, & \qquad (s = a) \\
				\frac{\mu_v^2 + \mu_a^2}{2} + \frac{\pi^2}{6} T^2 \,, & \qquad (s = \varepsilon)
			\end{array}
			\right.,  \label{eq:sigmaABa_ana} \\
			\sigma^{V}_{s}(0) &=& \frac{1}{2\pi^2} \left\{
			\begin{array}{ll}
				\mu_v \mu_a \,, & \qquad (s = v) \\
				\frac{\mu_v^2 + \mu_a^2}{2} + \frac{\pi^2}{6} T^2  \,, & \qquad (s = a) \\
				\frac{\mu_a}{3} \left( 3 \mu_v^2 + \mu_a^2  + \pi^2 T^2 \right)  \,, & \qquad (s = \varepsilon)
			\end{array}
			\right..     \label{eq:sigmaAV_ana}
		\end{eqnarray}
		In addition, all the conductivities related to the $w$ symmetry are vanishing when $M = 0$.  As a first step, before considering the massive case we have reproduced the results in the massless case. We present in figs.~\ref{fig:coryvM0}, \ref{fig:coryaM0} and \ref{fig:corytM0} some plots showing the comparison between the analytic results of eqs.~(\ref{eq:sigmaAB_ana})-(\ref{eq:sigmaAV_ana}) and the numerical results for $M = 0$.  It is clear that the numerics turns out to be consistent with the analytical formulas. 
		
		\subsection{Results of the conductivities vs. chemical potentials at finite $M$}
		\label{sec: result2}
		
		We present in this subsection the results for the conductivities as a function of the chemical potentials, for several values of $M/(\pi T)$. Here we will present only the $\mu_v$-dependence of the conductivities that induce the non-anomalous current $\vec{J}_w$, i.e. the conductivities appearing~in 
                \begin{equation}
                  \vec{J}_w = \sigma^B_{wv} \, \vec{B}_v + \sigma^B_{wa} \, \vec{B}_a + \sigma^B_{ww} \, \vec{B}_w + \sigma^V_{w} \, \vec{\Omega} \,. \label{eq:Jw}
                \end{equation}
We focus momentarily in this subsection on these conductivities because they allow us to provide an explicit answer to the question posed at the end of section~\ref{sec:constitutive}. The complete list of results for these and other conductivities is given in appendix~\ref{sec:App_finite_M}.

\begin{figure}[htbp]
			\centering
			\includegraphics[width=.4\textwidth]{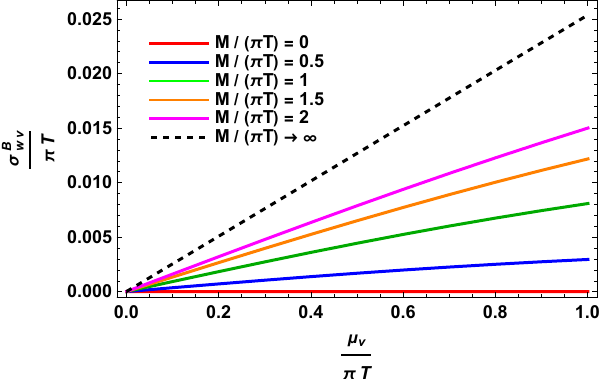}
			\qquad
			\includegraphics[width=.4\textwidth]{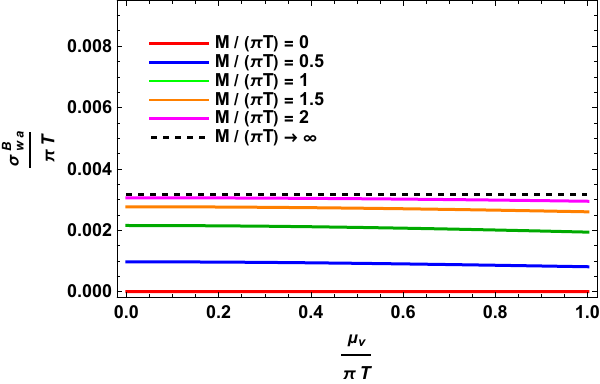}
			\qquad
			\includegraphics[width=.4\textwidth]{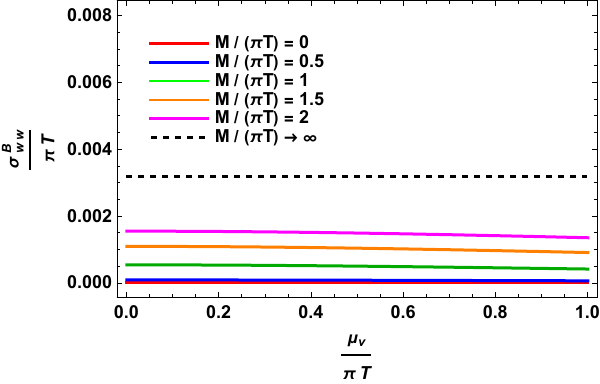}
			\qquad
			\includegraphics[width=.4\textwidth]{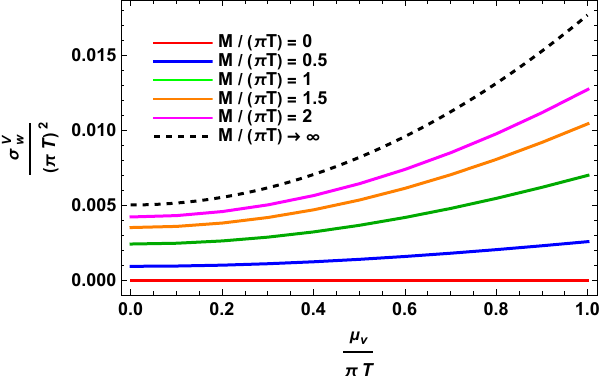}
			\caption{Plots of the conductivities $\sigma^B_{wv}$, $ \sigma^B_{wa}$,  $\sigma^B_{ww}$ and $\sigma^V_{w}$ as a function of to $\mu_v/(\pi T)$. We have set $\mu_a/(\pi T)=0.3$ and $\mu_w/(\pi T)=0.2$.  \label{fig:corywm}}
\end{figure}

                We display in fig.~\ref{fig:corywm} the results of the conductivities in eq.~(\ref{eq:Jw}) as a function of $\mu_v / (\pi T)$ in the range $0 \le M/(\pi T) \le 2$.  These results have been obtained by using the numerical method of sec.~\ref{subsec:numerical_method}. For comparison, we display also the results obtained in the large $M$ limit by using the analytical formulas that will be derived in sec.~\ref{subsec:large_M} and appendix~\ref{sec:App_large_M}. We observe that the conductivity $\sigma^{B}_{wv}$ exhibits a linear growth with $\mu_v/(\pi T)$, with a slope that increases as $M$ increases. On the other hand, the conductivities $\sigma^{B}_{wa}$ and $\sigma^{B}_{ww}$ remain constant with respect to $\mu_v/(\pi T)$, though their magnitudes shift upward with increasing $M$. Finally, we find for $\sigma^V_{w}$ a quadratic dependence on $\mu_v/(\pi T)$, with an increasing curvature as $M$ increases. Notice that all the conductivities show in general a pronounced dependence with the mass parameter $M$. 

		\subsection{Results of the conductivities vs. $M$ for fixed chemical potentials }
		\label{sec: result3}

		In this subsection we will show the results for the behaviour of the conductivities as a function of $M/(\pi T)$ in two different cases: {\it i)} fixed non-vanishing chemical potentials, and {\it ii)}~vanishing chemical potentials.

		\subsubsection{Fixed non-vanishing chemical potentials}

\begin{figure}[htbp]
			\centering
			\includegraphics[width=.47\textwidth]{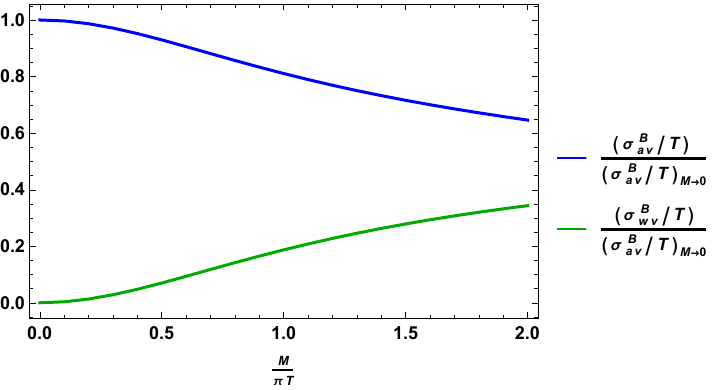}
			\qquad
			\includegraphics[width=.47\textwidth]{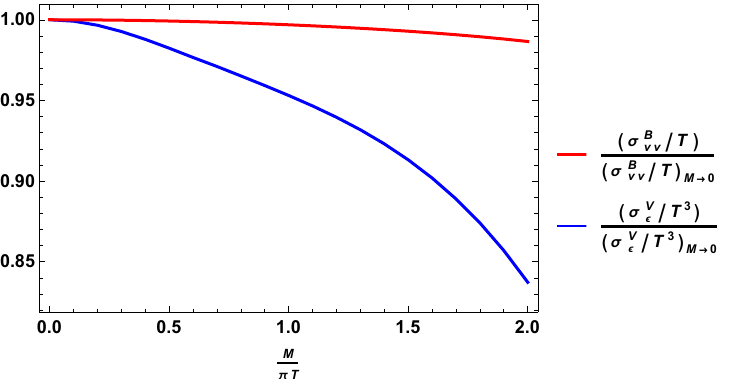}
			\caption{Plots of the chiral magnetic conductivities $\sigma^B_{s v}$ with $s \in \{ a,w \}$ (left panel) and $\sigma^B_{v v}$ along with $\sigma^V_{\varepsilon}$ (right panel), as a function of $M/(\pi T)$. We have set $\mu_v/(\pi T)=0.1$, $\mu_a/(\pi T)=0.3$ and $\mu_w/(\pi T) =0.2$.} \label{fig:comb1}
		\end{figure}

\begin{figure}[htbp]
			\centering
			\includegraphics[width=.47\textwidth]{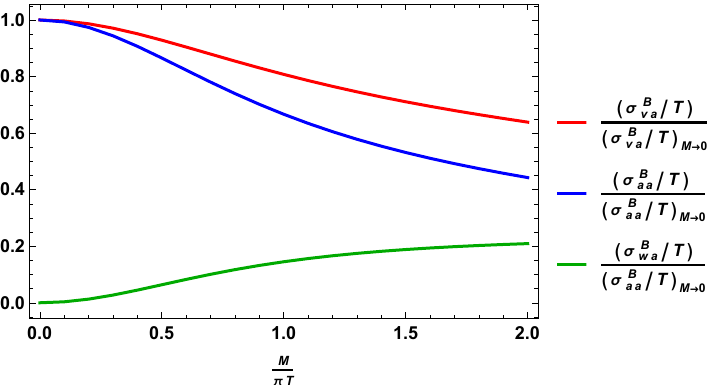}
			\qquad
			\includegraphics[width=.47\textwidth]{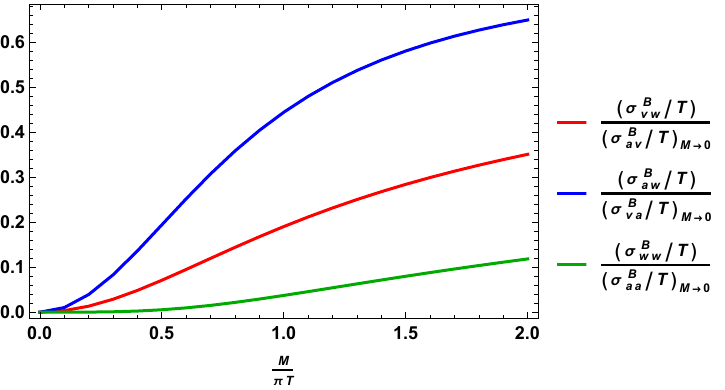}
			\caption{Plots of the chiral magnetic conductivities $\sigma^B_{s a}$ (left panel) and $\sigma^B_{s w}$ (right panel), where $s \in \{ v,a,w \}$, as a function of $M/(\pi T)$. See fig.~\ref{fig:comb1} for further details.} \label{fig:comb2}
		\end{figure}

\begin{figure}[htbp]
			\centering
			\includegraphics[width=.47\textwidth]{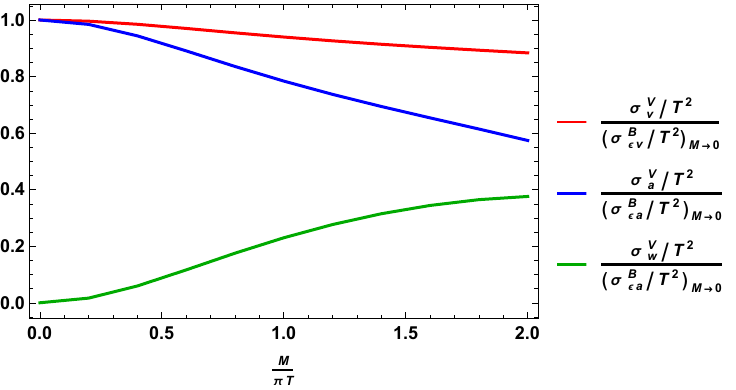}
			\caption{Plot of the chiral vortical conductivities $\sigma^V_s$, where $s \in \{ v, a, w \}$,  as a function of $M/(\pi T)$.  See fig.~\ref{fig:comb1} for further details.}
			\label{fig:comb3} 
		\end{figure}

		We will present below the results corresponding to fixing the chemical potentials as $\mu_v/(\pi T) = 0.1$,  $ \mu_a/(\pi T)=0.3$ and $\mu_w/(\pi T) = 0.2$.  We have computed the conductivities involving vector, axial, non-anomalous and energy currents as a function of the dimensionless mass parameter $M/(\pi T)$. We show the plots of the conductivities normalized by their $M\to0$ values when these are finite. In cases where the conductivities vanish in this limit, we normalized them by using some related non-vanishing conductivities from the same sector. The results are given in figs.~\ref{fig:comb1}, ~\ref{fig:comb2} and~\ref{fig:comb3}. We can see in these figures that the conductivities inducing the axial current $\vec{J}_a$ are monotonically suppressed, while the conductivities inducing the non-anomalous current $\vec{J}_w$ become enhanced when $M/(\pi T)$ increases. This indicates a redistribution of the magnetic response from the axial to the non-anomalous sector, while the vector sector is less affected.  In addition, all the chiral magnetic conductivities that control the response to a magnetic field in the non-anomalous sector ($\sigma^{B}_{vw}$, $\sigma^{B}_{aw}$, and $\sigma^{B}_{ww}$) increase with $M/(\pi T)$. The chiral vortical conductivities exhibit a complementary structure: the vector part ($\sigma^{V}_{v}$) is less affected by the change in $M / (\pi T)$, the axial part ($\sigma^{V}_{a}$) is suppressed, and the non-anomalous sector ($\sigma^{V}_{w}$) is amplified.

		\subsubsection{Zero chemical potentials}

\begin{figure}[htbp]
			\centering
			\includegraphics[width=.47\textwidth]{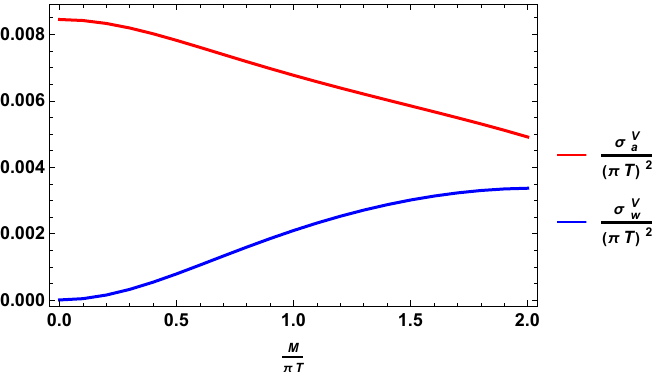}
			\caption{Plot of the conductivities $\sigma^V_a = \sigma^B_{\varepsilon a}$ and $\sigma^V_w = \sigma^B_{\varepsilon w}$ as a function of $M/(\pi T)$. We have set $\mu_v/(\pi T) = \mu_a/(\pi T) = \mu_w/(\pi T) = 0$ .\label{fig:corJaw00}}
		\end{figure}

		For completeness and clarity, we also present in fig.~\ref{fig:corJaw00} the results of the non-vanishing conductivities for zero chemical potentials $\left( \mu_v = \mu_a = \mu_w  = 0 \right)$ as a function of $M / (\pi T)$. We display $\sigma_a^V$ and $\sigma_w^V$, while the other two conductivities, $\sigma^B_{\varepsilon a}$ and $\sigma^B_{\varepsilon w}$, are obtained by using the relations discussed in section~\ref{sec:kubo_formulae}, cf. eq.~(\ref{eq:sigmaV_symmetric}). Other conductivities not shown in fig.~\ref{fig:corJaw00} are vanishing for zero chemical potentials.
		
		\subsubsection{Results in the large $M$ limit}
		\label{subsec:large_M}
		
		In the results presented above, we observe the following properties for both zero and finite chemical potentials
		\begin{equation}
			\sigma^B_{s w}(0) = \sigma^B_{w s}(0) = 0 = \sigma^V_{w}(0) \,, \qquad (s \in \{v, a, w, \varepsilon)) \,,
		\end{equation}
		i.e. all the conductivities related to the $w$ symmetry are vanishing in the massless case $(M = 0)$, which is the expected behavior in the absence of symmetry breaking. The values of the non-vanishing conductivities in the massless cases are given by eqs.~(\ref{eq:sigmaAB_ana})-(\ref{eq:sigmaAV_ana}).
		
		It would be relevant to study the behavior of the conductivities in the large $M$ limit $(M \to \infty)$. In order to derive analytical expressions for them, we will use the relation between the conductivities in the $(A,W)$ and $(A_+,A_-)$ basis. Let us consider for the moment the chiral magnetic conductivities as response of the vector magnetic field $\vec{B}_v$, which write
		\begin{equation}
			\sigma^B_{\pm v} \equiv \lim_{p_k \to 0}  \frac{i}{p_k} \langle  J_\pm J_v   \rangle = \sigma^B_{av} \pm \sigma^B_{wv}  \,,  \label{eq:sigmaB_pm}
		\end{equation}
		where we have used that $J_\pm = J_A \pm J_W$. For simplicity, we have omitted in eq.~(\ref{eq:sigmaB_pm}) the summation over Lorentz indices.   Inverting this expression, one finds
		\begin{equation}
			\sigma^B_{av} = \frac{1}{2} \left( \sigma^B_{+ v} + \sigma^B_{- v} \right) \,, \qquad \sigma^B_{wv } = \frac{1}{2} \left( \sigma^B_{+ v} - \sigma^B_{- v} \right) \,.  \label{eq:sigmaB_aw}
		\end{equation}
		While both symmetries $A_+$ and $A_-$ have CS interactions, the scalar field breaks only the $A_-$ symmetry. At $M=0$ there will be chiral conductivities for both $A_\pm$, but for large $M$, $A_-$ will be badly broken, and the chiral magnetic effect in the $J_{-}$ current goes to zero.~\footnote{See Ref.~\cite{Jimenez-Alba:2014iia} for a similar effect.} This implies the vanishing of $\sigma^B_{- v}$ in the limit $M \to \infty$, so that one has from eq.~(\ref{eq:sigmaB_aw}) the following relation
		\begin{equation}
			\sigma^B_{av}(\infty) = \sigma^B_{wv}(\infty) = \frac{1}{2} \sigma^B_{+v}(\infty)  \,.  \label{eq:sigmaB_av_wv}
		\end{equation}
		A similar analysis can be performed for the rest of the conductivities, and this leads to the following results
		\begin{equation}
			\left\{
			\begin{array}{l}
				\sigma^B_{av}(\infty) = \sigma^B_{wv}(\infty) = \frac{1}{2} \sigma^B_{+v}(\infty) \,, \\
				\sigma^B_{va}(\infty) = \sigma^B_{vw}(\infty) = \frac{1}{2} \sigma^B_{v+}(\infty) \,, \\
				\sigma^B_{\varepsilon a}(\infty) =  \sigma^B_{\varepsilon w}(\infty) = \frac{1}{2} \sigma^B_{\varepsilon +}(\infty) \,, \\
				\sigma^B_{a a}(\infty) =  \sigma^B_{w w}(\infty) = \sigma^B_{aw}(\infty) = \sigma^B_{wa}(\infty) = \frac{1}{4} \sigma^B_{++}(\infty)   \,, \\
				\sigma^V_a(\infty) = \sigma^V_w(\infty) = \frac{1}{2} \sigma^V_{+}(\infty)	 \,,
			\end{array}
			\right.         \label{eq:sigma_Minfty}  
		\end{equation}
		where for completeness we have included also the relation in eq.~(\ref{eq:sigmaB_av_wv}). In order to evaluate the asymptotic values of eq.~(\ref{eq:sigma_Minfty}), let us write the CS terms in the action in the $(A_+ , A_-)$ basis. It is given by
		\begin{eqnarray}
			\mathcal S_{\textrm{CS}} &=& \frac{1}{16\pi G} \int d^5x \sqrt{-g} \bigg[\frac{\kappa}{6} A_- \wedge F_A \wedge F_A + \frac{\kappa}{6} A_+ \wedge F_A \wedge F_A  \nonumber \\
			&&\hspace{3cm} + \frac{\kappa}{2} A_- \wedge  F_V \wedge F_V  + \frac{\kappa}{2} A_+ \wedge  F_V \wedge F_V \nonumber \\
			&&\hspace{3cm}+ \frac{\lambda}{2} A_- \wedge R \wedge R  + \frac{\lambda}{2} A_+ \wedge R \wedge R \bigg] \,,
		\end{eqnarray}
		where
		\begin{equation}
			F_A \wedge F_A = \frac{1}{4}  F_{A_-} \wedge F_{A_-} +  \frac{1}{4} F_{A_+} \wedge F_{A_+} + \frac{1}{2} F_{A_-} \wedge F_{A_+}  \,.
		\end{equation}
		Since $A_{-}$ is badly broken in the limit $M \to \infty$, one ends up in this limit with the CS contributions
		\begin{equation}
			\mathcal S_{\textrm{CS}} = \frac{1}{16\pi G} \int d^5x \sqrt{-g} \left[ \frac{\kappa}{24} A_+ \wedge F_{A_+} \wedge F_{A_+} + \frac{\kappa}{2} A_+ \wedge  F_V \wedge F_V + \frac{\lambda}{2} A_+ \wedge R \wedge R \right] \,.
		\end{equation}
		From a comparison with eq.~(\ref{act}), one deduces that the conductivities in the limit $M \to \infty$ can be obtained by making the replacement $\mu_a \to \mu_+ / 2$ in the massless case, eqs.~(\ref{eq:sigmaAB_ana})-(\ref{eq:sigmaAV_ana}). This leads to
		\begin{eqnarray}
			\sigma^{B}_{s v}(\infty) &=& \frac{1}{2\pi^2} \left\{
			\begin{array}{ll}
				\frac{\mu_+}{2} \,, & \qquad s = v \\
				\mu_v  \,, &  \qquad s = + \\
				\frac{\mu_v \mu_+}{2} \,, & \qquad s = \varepsilon
			\end{array}
			\right.,   \label{eq:sigmaAB_ana_M} \\
			\sigma^{B}_{s +}(\infty) &=& \frac{1}{2\pi^2} \left\{
			\begin{array}{ll}
				\mu_v \,, & \qquad s = v \\
				\frac{\mu_+}{2} \,, & \qquad s = + \\
				\frac{\mu_v^2 + \mu_+^2/4}{2} + \frac{\pi^2}{6} T^2 \,, & \qquad s = \varepsilon
			\end{array}
			\right.,  \label{eq:sigmaABa_ana_M} \\
			\sigma^{V}_{s}(\infty) &=& \frac{1}{2\pi^2} \left\{
			\begin{array}{ll}
				\frac{\mu_v \mu_+}{2} \,, & \qquad s = v \\
				\frac{\mu_v^2 + \mu_+^2/4}{2} + \frac{\pi^2}{6} T^2  \,, & \qquad s = + \\
				\frac{\mu_+}{6} \left( 3 \mu_v^2 + \mu_+^2/4 + \pi^2 T^2 \right)  \,, & \qquad s = \varepsilon
			\end{array}
			\right..     \label{eq:sigmaAV_ana_M}
		\end{eqnarray}
		In particular, by using eqs.~(\ref{eq:sigmaAB_ana}), (\ref{eq:sigmaB_av_wv}) and (\ref{eq:sigmaAB_ana_M}) one finds that
		\begin{equation}
			\sigma^B_{av}(\infty) = \sigma^B_{wv}(\infty) = \frac{1}{2} \sigma^B_{+v}(\infty) = \frac{\mu_v}{4\pi^2} = \frac{1}{2} \sigma^B_{av}(0) \,. \label{eq:sigma_av_Minfty}
		\end{equation}
		The expressions for the other conductivities in eq.~(\ref{eq:sigma_Minfty}) can be read off from eqs.~(\ref{eq:sigmaAB_ana_M})-(\ref{eq:sigmaAV_ana_M}). The results for all the conductivities in the large $M$ limit and in the basis $(V,A,W)$ are collected in appendix~\ref{sec:App_large_M}. Notice that the values of the conductivities in this limit are not directly related in general to their values in the massless case due to their dependence on $\mu_+ \equiv \mu_a + \mu_w$, except for the case presented in eq.~(\ref{eq:sigma_av_Minfty}). In addition, it turns out that in general all the conductivities have different values at $M = 0$ and at $M \to \infty$, thus implying that all of them have a dependence on the symmetry breaking parameter $M$. However, there is a single particular non-trivial situation in which some of the conductivities have the same values in both limits, i.e.  $\sigma^B_{vv}$, $\sigma^B_{\varepsilon v}$, $\sigma^V_v$ and $\sigma^V_\varepsilon$ for $\mu_a = \mu_w$.~\footnote{The trivial situation in which this happens corresponds to $T = 0$ and/or $\mu_s = 0 \;\;  (s \in \{v,a,w\})$.} We have checked that if $\mu_a = \mu_w$, this is not just an asymptotic property but in fact these four conductivities turn out to be independent of the mass parameter~$M$.

\begin{figure}[htbp]
			\centering
			\includegraphics[width=.47\textwidth]{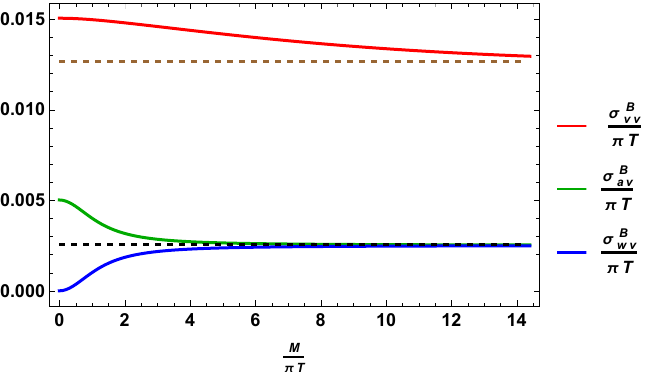}
			\qquad
			\includegraphics[width=.47\textwidth]{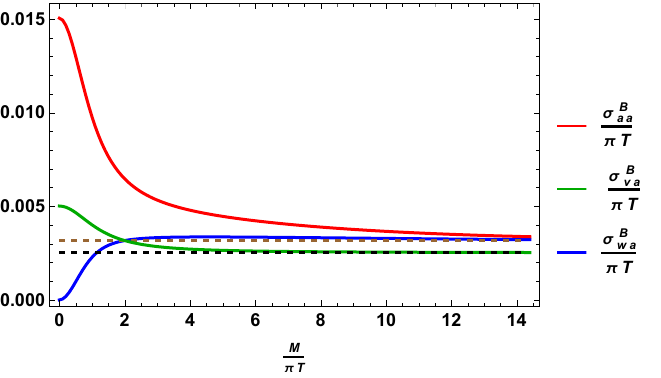}
			\caption{Plots of the chiral magnetic conductivities $\sigma^B_{s v}$ (left panel) and $\sigma^B_{s a}$ (right panel), where $s \in \{ v,a,w \}$, as a function of $M/(\pi T)$ in the probe limit. The dashed horizontal lines correspond to the analytic results given by eqs.~(\ref{eq:App_sigmaAB_ana_M}) and (\ref{eq:App_sigmaABa_ana_M}). We have set $\mu_v/(\pi T)=0.1$,  $\mu_a/(\pi T)=0.3$ and $\mu_w/(\pi T) = 0.2$.}  \label{fig:probe_va}
		\end{figure}

\begin{figure}[htbp]
			\centering
			\includegraphics[width=.47\textwidth]{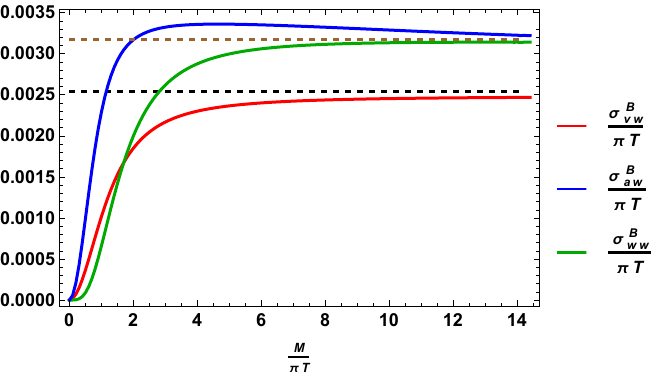}
			\caption{Plot of the chiral magnetic conductivities $\sigma^B_{s w}$, where $s \in \{ v,a,w \}$, as a function of $M/(\pi T)$ in the probe limit.  The dashed horizontal lines correspond to the analytic results given by eq.~(\ref{eq:App_sigmaABw_ana_M}). See fig.~\ref{fig:probe_va} for further details.} \label{fig:probe_w}
		\end{figure}

		We see in figs.~\ref{fig:comb1}, \ref{fig:comb2}, \ref{fig:comb3} and \ref{fig:corJaw00} that the tendency of the numerical results is to confirm the relations of eq.~(\ref{eq:sigma_Minfty}). A more robust check demands computing the conductivities for higher values of $M / (\pi T)$. However, this problem turns out to be difficult when considering the full backreaction of the metric due to the great increase of the numerical error for larger values of~$M$. A way of overcoming this difficulty is to work in the probe limit, as we expect that the computation with no backreaction leads to a good approximation in the limit $M/(\pi T) \gg 1$ for conductivities whose results just depend on the chemical potentials and not on the temperature, a property that is confirmed by the numerics: this is the case of conductivities related to charged transport induced by magnetic fields, but not those related to the vortical effect or the energy transport sector which demand a computation with full backreaction of the metric. We display in figs.~\ref{fig:probe_va} and \ref{fig:probe_w} the numerical results in the probe limit for the conductivities in the charged sector in the range $0 \le M/(\pi T) \le 15$, together with the corresponding analytical results in the large $M$ limit (cf. appendix~\ref{sec:App_large_M}). Note that the agreement between numerics and analytical formulas is excellent for $M/(\pi T) \gtrsim 10$.
		

		\section{Discussion and outlook}
		\label{sec:discussion}
		
		In this work, we have studied the effect of explicit symmetry breaking in anomaly-induced transport by using a holographic framework that includes both pure gauge and mixed gauge-gravitational Chern-Simons terms. Our findings show that when symmetry is explicitly broken, anomalies can have a considerable impact not only on the anomalous currents but also affect the non-anomalous sectors as well. 
		
		Firstly, we have performed a study in the massless case $(M=0)$ which corresponds to the symmetric phase. Our numerical results reproduce in this case the known analytic formulas, thus confirming the reliability of our approach. Next, we have studied the problem at finite $M$, and observed that all the conductivities have a dependence on the symmetry breaking parameter, $M$, and on the chemical potentials ($\mu_v$, $\mu_a$, $\mu_w$). We have characterized these dependencies both numerically at finite $M$, and analytically in the asymptotic limit $M \to \infty$. Our results also reveal that even at vanishing chemical potentials, certain transport coefficients ($\sigma^V_a$, $\sigma^V_w, \sigma^B_{\varepsilon a}$, $\sigma^B_{\varepsilon w}$) are non-vanishing at finite temperature, implying that symmetry breaking alone is sufficient to induce non-trivial anomalous responses. These findings highlight the role of the mixed gauge–gravitational anomaly in shaping holographic transport properties and suggest that anomaly-induced contributions to non-anomalous currents could have observable consequences in strongly coupled systems where symmetries are not exact.

		In summary, our results reveal how symmetry breaking terms interact explicitly with anomaly contributions to modify the response to anomalous transport. One of the most interesting results of this work is that once the symmetry is broken by the scalar field $\phi$, even the non-anomalous current $\vec{J}_w$ shows response akin to the chiral vortical and chiral magnetic effect. It would be interesting to study the analogous phenomenon in a weak coupling field theoretical setup. One might also think of possible application in condensed matter physics. Weyl semi-metals typically come with a multiple of Weyl nodes (valleys) in the Brillouin zone~\cite{Basar:2013iaa,Landsteiner:2015lsa}. One can speculate that they feature not only axial symmetries by also some non-anomalous accidental valley symmetries and corresponding valley currents. It would be fascinating to see a similar non-anomalous CME in such valley currents in certain phases in which parts of valley symmetries are broken.

		\begin{acknowledgments}
			E.M. thanks Grigory E. Volovik for useful correspondence. A.T. would like to thank his supervisor Shubhrangshu Ghosh for his valuable suggestions and help throughout the work. A.T. acknowledges financial support by the Ministry of Tribal Affairs, Government of India, through the National Fellowship for Scheduled Tribes (NFST). The works of N.R. and E.M. are supported by the ``Proyectos de Investigaci\'on Precompetitivos'' Program of the Plan Propio de Investigaci\'on of the University of Granada under grant PP2025PP-18. N.R. and E.M. acknowledge financial support during the initial stages of this work by the Ram\'on y Cajal Program under Grant RYC-2016-20678 funded by MCIN/AEI/10.13039/50110\-0011033 and by ``FSE Investing in your future'', and by the FEDER/Junta de Andaluc\'{\i}a-Consejer\'{\i}a de Econom\'{\i}a y Conocimiento 2014–2020 Operational Program under Grant A-FQM-178-UGR18. The work of K.L. is partially supported by CEX2020-001007-S and PID2021-123017NB-I00, PID2021-127726NB-I00,  PID2024-156043NB-I00 funded by MCIN/AEI/10.13039/501100011033 and ERDF ``A way of making Europe''. 
		\end{acknowledgments}
		
		\begin{appendices}

			\section{Equations of motion}
			\label{sec:appendix_EoM}
			
			We present in this appendix the explicit expressions of the equations of motion for both the background and the fluctuations, for the metric and for the gauge fields. We are assuming that the momentum $\vec{p}$ is directed along the $z$-axis, i.e. $\vec{p} = p \hat{z}$.
			
			\subsection{Equations of motion of the background}
			\label{subsec:appendix_background}
			
			The equations of motion of the background are given by
			\begin{eqnarray}
				0 &=& V_t^{\prime\prime} - \frac{\chi'}{2} V_t^\prime \,, \label{eq:Vt} \\
				0 &=&  A_{t+}^{\prime\prime} - \frac{\chi'}{2} A_{t+}' \,, \label{eq:Apt} \\
				0 &=& A_{t-}^{\prime\prime} - \frac{\chi'}{2} A_{t-}' - \frac{e^{\chi}}{ u^2 f} \phi^2 A_{t-}  \,,  \label{eq:Amt}  \\ 
				0 &=&  \phi^{\prime\prime} - \left( \frac{1}{u} + \frac{\chi^\prime}{2} - \frac{f^\prime}{f} \right) \phi^\prime  + \frac{e^{\chi}}{4 u^2 f^2} \left( 3 f + u A_{t-}^2 \right)  \phi  \,,  \label{eq:phi}  \\
				0 &=& \phi^{\prime \, 2} + \frac{3 \chi^\prime}{4u} + \frac{e^{\chi} }{ 4 u f^2} \phi^2   A_{t-}^2   \,, \label{eq:f}  \\
				0 &=& f^\prime - \frac{f}{2u} (4 + u \chi^\prime)  + \frac{e^{\chi}}{2u} ( 4 + \phi^2 ) - \frac{u^2}{6} \left( 2 V_t^{\prime \, 2}  + A_{t-}^{\prime \, 2} + A_{t+}^{\prime \, 2}  \right) \,, \label{eq:chi}
			\end{eqnarray}
			where the primes stand for derivatives with respect to $u$. These equations follow from: {\it i)}~the Maxwell's equations of the gauge fields, eqs.~(\ref{eq:Vt})-(\ref{eq:Amt}); {\it ii)}~the Klein-Gordon equation of the scalar field, eq.~(\ref{eq:phi}); and {\it iii)}~the Einstein's equations, eqs.~(\ref{eq:f})-(\ref{eq:chi}). In this appendix we are using the notation $A_{t\pm} \equiv (A_\pm)_{t}$.

			\subsection{Equations of motion of the fluctuations}
			\label{subsec:appendix_fluctuations}
			
			The equations of motion of the fluctuations are given by
			\begin{eqnarray}
				0 &=& v_x'' +  \left( \frac{f'}{f} - \frac{\chi'}{2} \right) v_x' + \frac{ V_t'}{f} h_{t}^{x \, \prime} - i p \kappa \frac{2 e^{\chi/2}}{f} \left[
				(A_{t-}'+ A_{t+}') v_y 
				+V_t' \left((a_{-})_y
				+ (a_{+})_y\right)
				\right]  \,, \label{eq:vx} \\
				0 &=& v_y'' +  \left( \frac{f'}{f} - \frac{\chi'}{2} \right) v_y' +  \frac{V_t'}{f} h_{t}^{y \, \prime}  + i p \kappa \frac{2 e^{\chi/2}}{f} \left[(A_{t-}'+ A_{t+}') v_x 
				+V_t' \left((a_{-})_x
				+ (a_{+})_x\right)
				\right]  \,, \label{eq:vy} \\
				0 &=& (a_{-})_x'' + \left(\frac{f'}{f} - \frac{\chi'}{2}\right) (a_{-})_x'- \frac{e^{\chi}\phi^2}{u^2 f} (a_{-})_x +\frac{A_{t-}'}{f}h_{t}^{x \, \prime}  \label{eq:a1x}  \\
				\hspace{-0.7cm}	&&-i \frac{p}{f} \left[ \kappa \,  e^{\chi/2} \left( \left(A_{t-}'+A_{t+}'\right) \left((a_{-})_y + (a_{+})_y \right) + 4 V_t' v_y\right) - 4 \lambda \, e^{-\chi/2} \left( \left(1 - u\chi'\right) f^\prime +2uf^{\prime\prime}  \right) h_{t}^{y \, \prime }  \right] \,, \nonumber  \\
				0 &=& (a_{-})_y'' + \left(\frac{f'}{f} - \frac{\chi'}{2}\right) (a_{-})_y'- \frac{e^{\chi}\phi^2}{u^2 f} (a_{-})_y +\frac{A_{t-}'}{f}h_{t}^{y \, \prime}  \label{eq:a1y} \\
				\hspace{-0.7cm}	&&+i \frac{p}{f} \left[ \kappa \, e^{\chi/2} \left( \left(A_{t-}'+A_{t+}'\right) \left((a_{-})_x + (a_{+})_x \right) + 4 V_t' v_x\right) - 4 \lambda \, e^{-\chi/2}  \left( \left(1 - u\chi'\right) f^\prime +2uf^{\prime\prime} \right)  h_{t}^{x \, \prime} \right] \,, \nonumber   \\
				0 &=& (a_{+})_x'' + \left(\frac{f'}{f} - \frac{\chi'}{2}\right) (a_{+})_x' + \frac{A_{t+}'}{f}h_{t}^{x \, \prime}  \label{eq:a2x}  \\
				\hspace{-0.7cm}	&& -i \frac{p}{f} \left[\kappa \, e^{\chi/2} \left( \left(A_{t-}'+A_{t+}'\right) \left((a_{-})_y + (a_{+})_y\right) + 4 V_t' v_y\right) - 4 \lambda  \, e^{-\chi/2}   \left( \left(1 - u\chi'\right) f^\prime +2uf^{\prime\prime}  \right) h_{t}^{y \, \prime} \right] \,, \nonumber \\
				0 &=& (a_{+})_y'' + \left(\frac{f'}{f} - \frac{\chi'}{2}\right) (a_{+})_y' +\frac{A_{t+}'}{f}h_{t}^{y \, \prime}  \label{eq:a2y}   \\
				\hspace{-0.7cm}	&&+i \frac{p}{f}\left[\kappa \, e^{\chi/2} \left(  \left(A_{t-}'+A_{t+}'\right) \left((a_{-})_x + (a_{+})_x \right) + 4 V_t'  v_x\right) - 4 \lambda \, e^{-\chi/2}  \left( \left(1 - u\chi'\right) f^\prime +2uf^{\prime\prime}  \right)  h_{t}^{x \, \prime} \right] \,, \nonumber  \\
				0 &=& h_{t}^{x \, \prime\prime} -\frac{\left(2+u\chi'\right)}{2u}h_{t}^{x \, \prime} + uV_t'v_x' \nonumber \\
				&& +\frac{u}{2}\left(A_{t-}' (a_{-})_x'+A_{t+}' (a_{+})_x'\right)  +\frac{e^\chi A_{t-}\phi^2}{2uf}(a_{-})_x - 2i p \lambda u \, e^{-\chi/2} \,\Gamma_y  \,,   \label{eq:hxt}  \\
				0 &=& h_{t}^{y \, \prime\prime} -\frac{\left(2+u\chi'\right)}{2u}h_{t}^{y \, \prime} +  u V_t'v_y' \nonumber \\
				&&+\frac{u}{2}\left(A_{t-}' (a_{-})_y'+A_{t+}' (a_{+})_y'\right) +\frac{e^\chi A_{t-}\phi^2}{2uf}(a_{-})_y + 2i p \lambda  u \, e^{-\chi/2}  \,\Gamma_x \,, \label{eq:hyt}
			\end{eqnarray}
			where the terms $\Gamma_j \; (j = x, y)$  write
			\begin{eqnarray}
				\Gamma_j  &=& 2 u \,  (A_{t-}^\prime + A_{t+}^\prime ) \, h_{t}^{j \, \prime\prime}  + 2  \left[
				(1 - u \chi^\prime )  ( A_{t-}^\prime + A_{t+}^\prime)  +  u (A_{t-}'' + A_{t+}'')  \right] h_{t}^{j \, \prime}	\nonumber  \\
				\hspace{-0.7cm}	&& -\left[ (1 - u \chi') f^\prime + 2u f'' \right] ( (a_{-})_j^\prime  + (a_{+})_j^\prime )  \nonumber \\
				\hspace{-0.7cm} && + \left[  (2 \chi^\prime - u \chi^{\prime \, 2} + u \chi^{\prime\prime} )  f^\prime - 3 (1 - u \chi^\prime) f^{\prime\prime} - 2 u f^{(3)} \right]   ( (a_{-})_j + (a_+)_j )   \,. \hspace{0.5cm} \label{eq:Gamma_j}
			\end{eqnarray}
			In these expressions we have set $\omega = 0$, and have neglected non-anomalous terms of~$\mathcal O(p^2)$ and anomalous terms of~$\mathcal O(p^3)$.

			\section{Conductivities at finite $M$}
			\label{sec:App_finite_M}
			
			We present in this appendix all the numerical results for the conductivities appearing in eqs.~(\ref{eq:Jv_const})-(\ref{eq:Je_const}) as a function of $\mu_v / (\pi T)$, $\mu_a / (\pi T)$ and $\mu_w / (\pi T)$, for several values of $M / (\pi T)$. When presenting the results, we will classify them in four categories corresponding to the conductivities controlling the response to: {\it i)} $\vec{B}_v$, {\it ii)} $\vec{B}_a$, {\it iii)} $\vec{B}_w$, and {\it iv)} $\vec{\Omega}$; in this order. Unless otherwise stated, we will use as fixed values of the chemical potentials the following ones: $\mu_v / (\pi T) = 0.1$, $\mu_a / (\pi T) = 0.3$ and $\mu_w / (\pi T) = 0.2$; and we will consider three different cases: \textbf{Case 1:} changing $\mu_v / (\pi T)$, \textbf{Case 2:} changing $\mu_a / (\pi T)$, and \textbf{Case 3:} changing $\mu_w / (\pi T)$. We will show the results obtained by using the numerical method of sec.~\ref{subsec:numerical_method} in the range $0 \le M/(\pi T) \le 2$. For comparison, we will also display the results obtained in the large $M$ limit by using the analytical formulas that will be derived in sec.~\ref{subsec:large_M} and appendix~\ref{sec:App_large_M}.
			
				\begin{figure}[t]
				\centering
				\includegraphics[width=.4\textwidth]{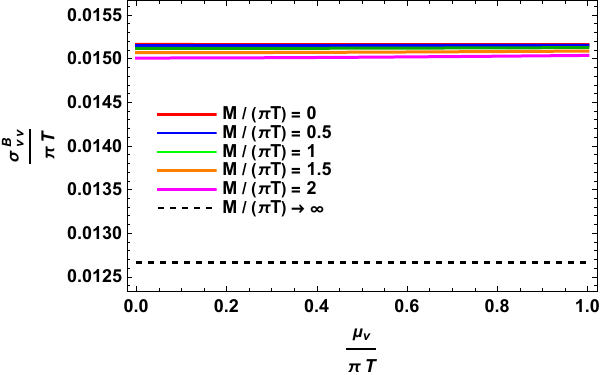}
				\qquad
				\includegraphics[width=.4\textwidth]{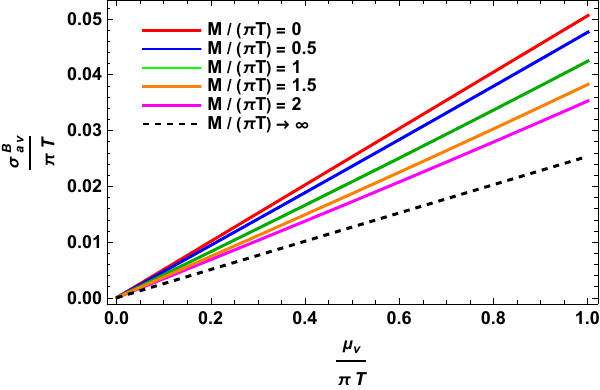}
				\qquad
				\includegraphics[width=.4\textwidth]{wvv}
				\qquad
				\includegraphics[width=.4\textwidth]{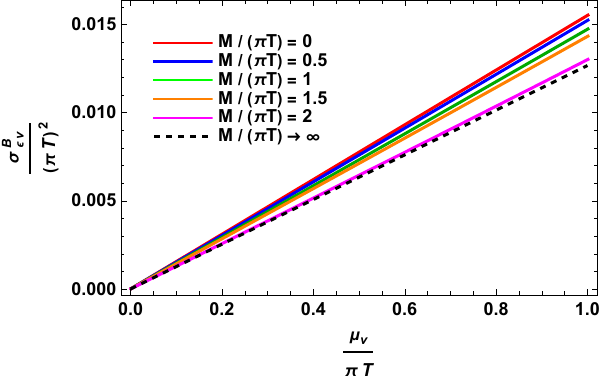}
				\caption{Case 1: Plots of the conductivities $\sigma^B_{s v}$, where $s \in \{ v, a, w, \varepsilon \}$, as a function of $\mu_v/(\pi T)$. We have set $\mu_a/(\pi T)=0.3$ and $\mu_w/(\pi T)=0.2$.}
				\label{fig:coryv}
			\end{figure}

			\subsection{Conductivities controlling the response to $\vec{B}_v$}
			\label{subsec:Bv}

			We begin with the case where the source term of $\vec{B}_v$ is switched on while the remaining sources are set to zero. This leads to 12 plots: four conductivities $(\sigma^B_{vv}, \sigma^B_{av}, \sigma^B_{wv}, \sigma^B_{\varepsilon v})$ with variation of three different chemical potentials $(\mu_v, \mu_a, \mu_w)$. We present below the results distributed according to the three different cases mentioned above.
			
\begin{figure}[htbp]
				\centering
				\includegraphics[width=.4\textwidth]{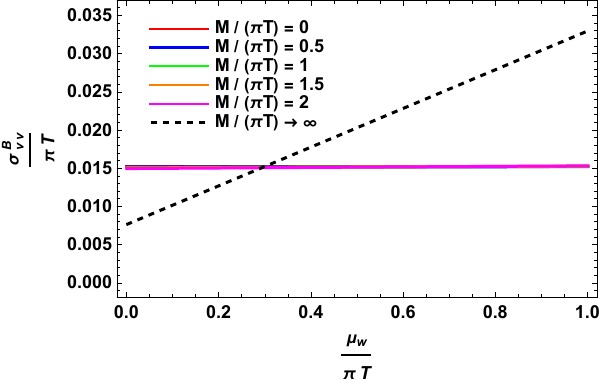}
				\qquad
				\includegraphics[width=.4\textwidth]{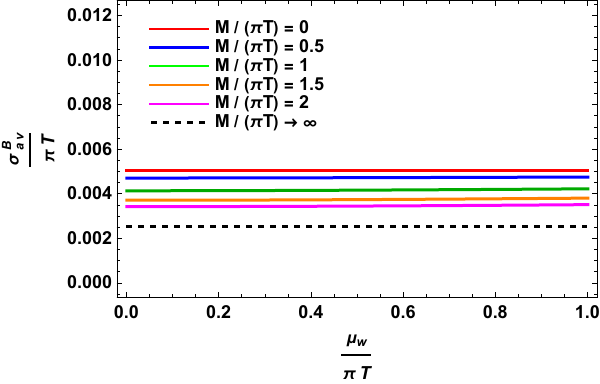}
				\qquad
				\includegraphics[width=.4\textwidth]{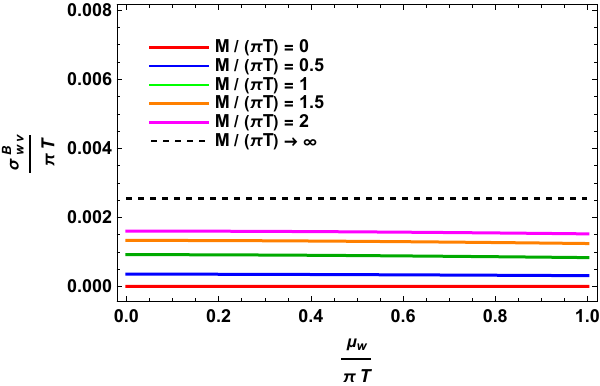}
				\qquad
				\includegraphics[width=.4\textwidth]{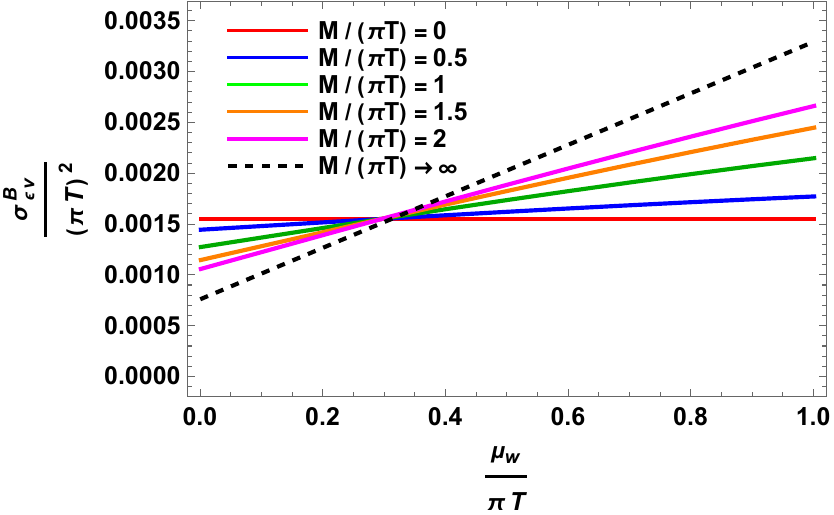}
				\caption{Case 3: Plots of the conductivities $\sigma^B_{s v}$, where $s \in \{ v, a, w, \varepsilon \}$, as a function of $\mu_w/(\pi T)$. We have set  $\mu_v/(\pi T)=0.1 $ and $\mu_a/(\pi T)=0.3$. \label{fig:coryv2}}
			\end{figure}

			\begin{itemize}
				\item \textbf{Case 1:} The results are displayed in fig.\;\ref{fig:coryv}. We can see that  $\sigma_{av}^B$,  $\sigma_{wv}^B$ and $\sigma_{\varepsilon v}^B$ grow linearly with $\mu_v / (\pi T)$. With the increase $M$  the slopes of  $\sigma_{av}^B$ and $\sigma_{\varepsilon v}^B$ decrease, while the slope of $\sigma_{wv}^B$ increases. The conductivity $\sigma_{v v}^B$ turns out to be independent of $\mu_v / (\pi T)$ independently of the value of $M$, exactly as in the absence of symmetry breaking, cf. eq.~(\ref{eq:sigmaAB_ana}) \cite{Landsteiner:2013aba}.
				
				\item \textbf{Case 2:} The results are displayed in fig.\;\ref{fig:coryv1}. We can see that the conductivities $\sigma^{B}_{vv}$ and $\sigma^{B}_{\varepsilon v}$ grow linearly with $\mu_{a}/(\pi T)$. The slope of $\sigma^B_{vv}$ is almost unaffected by $M$ in the range $0 \le M / (\pi T) \le 2$, although a noticeable change appears for larger values of $M$ (this can be inferred from the left panel of fig.~\ref{fig:probe_va}). The remaining conductivities, $\sigma^B_{av}$ and $\sigma^B_{wv}$, turn out to be independent of $\mu_a / (\pi T)$ independently of the value of $M$, as it was the case for $M = 0$.

				\begin{figure}[htbp]
					\centering
					\includegraphics[width=.4\textwidth]{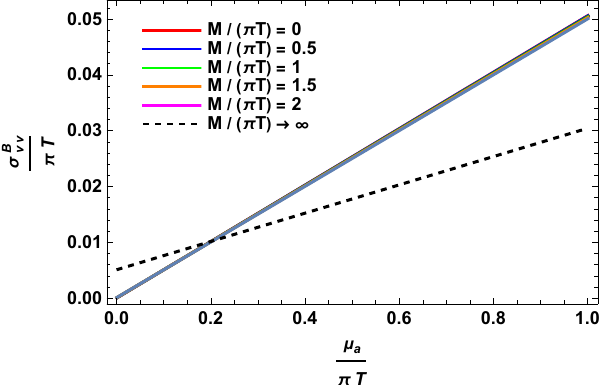}
					\qquad
					\includegraphics[width=.4\textwidth]{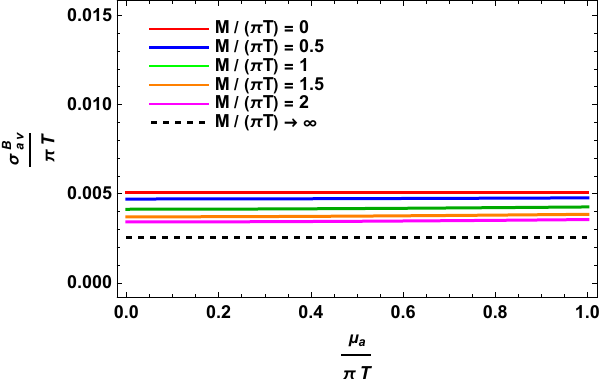}
					\qquad
					\includegraphics[width=.4\textwidth]{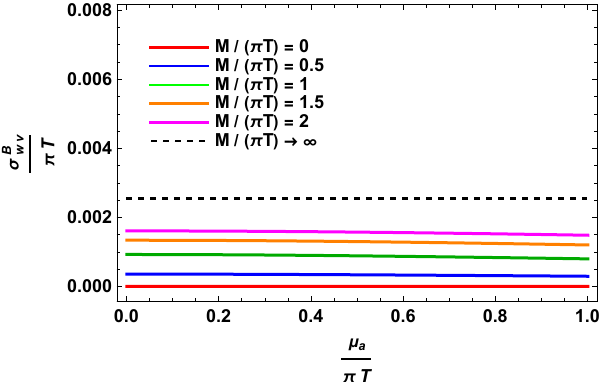}
					\qquad
					\includegraphics[width=.4\textwidth]{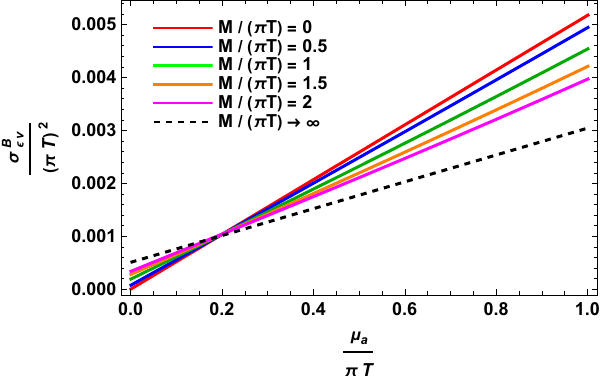}
					\caption{Case 2: Plots of the conductivities $\sigma^B_{s v}$, where $s \in \{ v, a, w, \varepsilon \}$, as a function of $\mu_a/(\pi T)$. We have set $\mu_v/(\pi T)=0.1$ and $\mu_w/(\pi T)=0.2$. \label{fig:coryv1}}
				\end{figure}

				\item \textbf{Case 3:} The results are displayed in fig.~\ref{fig:coryv2}. We can see that $\sigma^{B}_{\varepsilon v}$ grows linearly with $\mu_w / (\pi T)$, while $\sigma^{B}_{av}$ and $\sigma^{B}_{wv}$ remain constant and independent of $\mu_{w}/(\pi T)$. $\sigma^B_{vv}$ is almost independent of $\mu_w / (\pi T)$ in the range $0 \le M / (\pi T) \le 2$, although a non-vanishing slope appears for larger values of $M$, a behaviour that mirrors the trends observed in Case 2.  
			\end{itemize}

			\subsection{Conductivities controlling the response to $\vec{B}_a$}
			\label{subsec:Ba}

\begin{figure}[htbp]
				\centering
				\includegraphics[width=.4\textwidth]{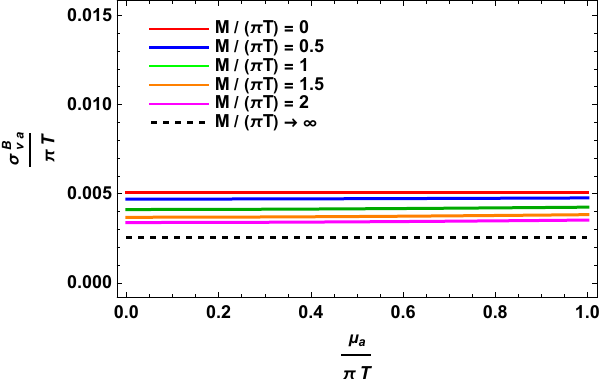}
				\qquad
				\includegraphics[width=.4\textwidth]{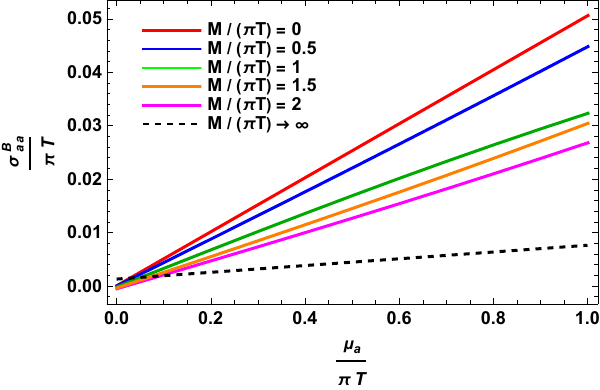}
				\qquad
				\includegraphics[width=.4\textwidth]{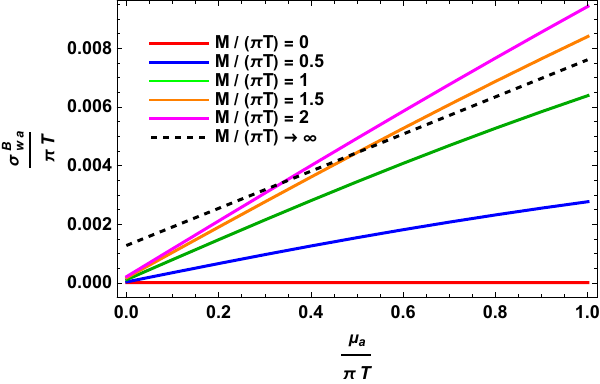}
				\qquad
				\includegraphics[width=.4\textwidth]{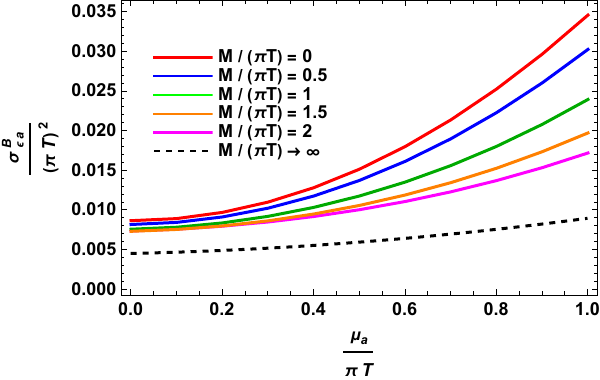}
				\caption{Case 2: Plots of the conductivities $\sigma^B_{s a}$, where $s \in \{ v, a, w, \varepsilon \}$, as a function of $\mu_a/(\pi T)$. We have set $\mu_v/(\pi T)=0.1$ and $\mu_w/(\pi T)=0.2$. \label{fig:corya1}}
			\end{figure}

\begin{figure}[htbp]
				\centering
				\includegraphics[width=.4\textwidth]{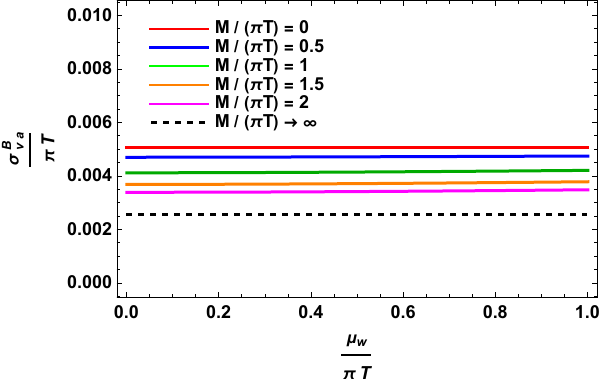}
				\qquad
				\includegraphics[width=.4\textwidth]{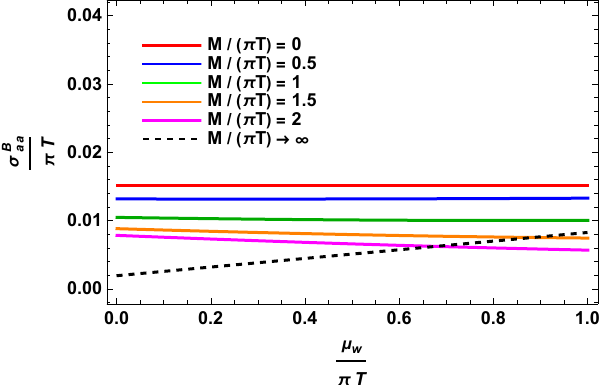}
				\qquad
				\includegraphics[width=.4\textwidth]{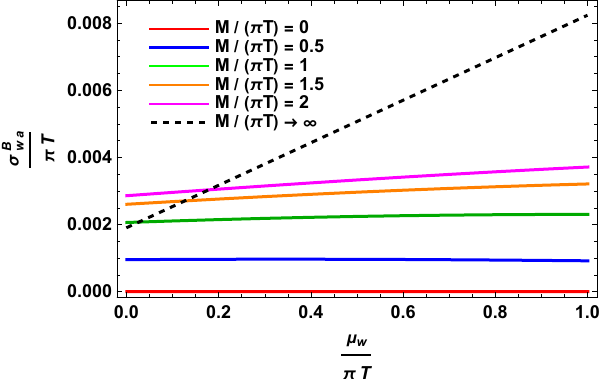}
				\qquad
				\includegraphics[width=.4\textwidth]{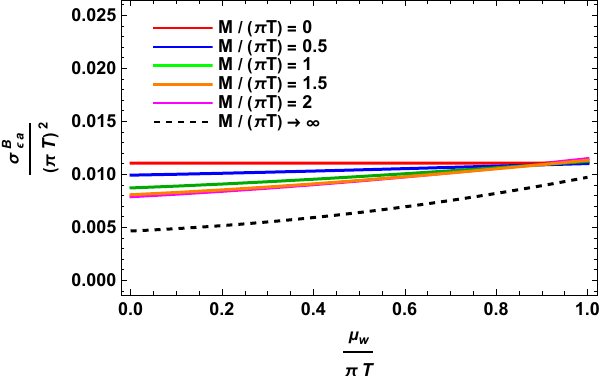}
				\caption{Case 3: Plots of the conductivities $\sigma^B_{s a}$, where $s \in \{ v, a, w, \varepsilon \}$, as a function of $\mu_w/(\pi T)$. We have set  $\mu_v/(\pi T)=0.1$ and $\mu_a/(\pi T)=0.3$. \label{fig:corya2}}
			\end{figure}

			Similar to the previous case, the source term of $\vec{B}_a$ is now switched on and the remaining sources are set to zero.
			\begin{figure}[t]
				\centering
				\includegraphics[width=.4\textwidth]{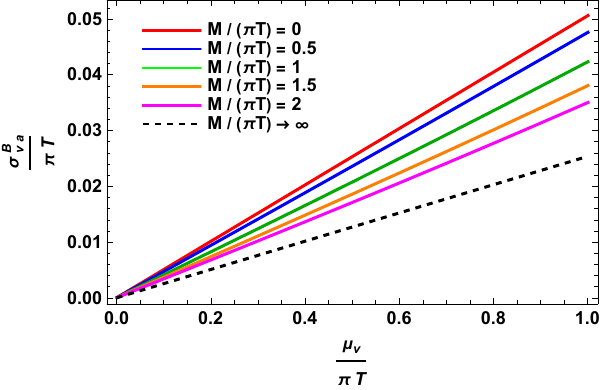}
				\qquad
				\includegraphics[width=.4\textwidth]{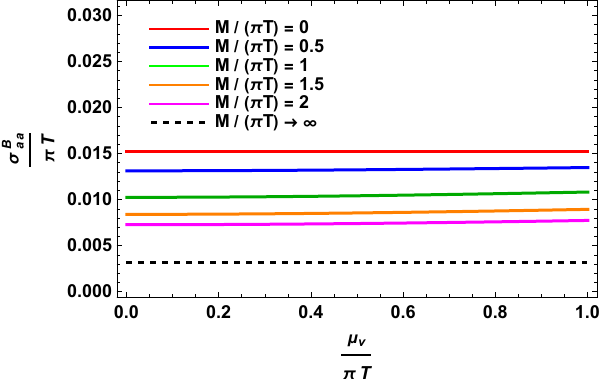}
				\qquad
				\includegraphics[width=.4\textwidth]{waa}
				\qquad
				\includegraphics[width=.4\textwidth]{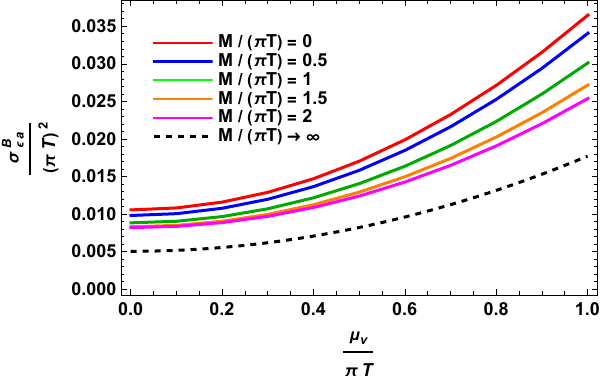}
				\caption{Case 1: Plots of the conductivities $\sigma^B_{s a}$, where $s \in \{ v, a, w, \varepsilon \}$, as a function of $\mu_v/(\pi T)$. We have set $\mu_a/(\pi T)=0.3$ and $\mu_w/(\pi T)=0.2$.\label{fig:corya}}
			\end{figure}
			\begin{itemize}
				\item \textbf{Case 1:} The results are displayed in fig.~\ref{fig:corya}. We observe that $\sigma^{B}_{va}$ increases linearly with $\mu_v/(\pi T)$, showing a decreasing slope as $M$ increases.  On the other hand, the conductivities $\sigma^B_{aa}$ and $\sigma^B_{wa}$ remain independent of $\mu_v/(\pi T)$, but their overall values decrease and increase, respectively, with increasing $M$. In contrast, $\sigma^{B}_{\varepsilon a}$ displays a quadratic dependence on $\mu_v/(\pi T)$, with a smaller curvature for higher values of $M$.  
				
				\item \textbf{Case 2:} The results are displayed in fig.~\ref{fig:corya1}. $\sigma^B_{va}$ remains constant and independent of $\mu_a/(\pi T)$, but its overall value decreases with increasing $M$.  The conductivities $\sigma^B_{aa}$ and $\sigma^B_{w a}$ exhibit a linear growth with $\mu_{a}/(\pi T)$, with slopes that decrease and increase, respectively, as $M$ increases in the range $0 \le M / (\pi T) \le 2$. This trend continues in the case of $\sigma^{B}_{aa}$ for higher values of $M$, but it becomes reverted in the case of $\sigma^{B}_{wa}$, as this conductivity exhibits a smaller slope for very large values of $M$. On this respect, let us comment that the intersection of the curves of $\sigma^B_{wa}$ for $M / (\pi T) = 2$ and $M / (\pi T) \to \infty$, happening at $\mu_a / (\pi T) \simeq 0.3$, can be seen also in fig.~\ref{fig:probe_w}, i.e. the conductivity $\sigma^B_{wa} = \sigma^B_{aw}$ displayed as a function of $M / (\pi T)$ and for fixed values of the chemical potentials, has the same values at $M / (\pi T) = 2$ and in the limit $M \to \infty$.  Finally, $\sigma^{B}_{\varepsilon a}$ displays a quadratic dependence on $\mu_a/(\pi T)$, with a smaller curvature for higher values of $M$.
				
				\item \textbf{Case 3:} The results are displayed in fig.~\ref{fig:corya2}. We observe that the conductivity $\sigma^{B}_{va}$ remains constant with respect to $\mu_w/(\pi T)$, while its overall value decreases as $M$ increases. The conductivities $\sigma^B_{aa}$ and $\sigma^{B}_{wa}$ show a linear behavior with $\mu_{w} / (\pi T)$, but their slopes turn out to be noticeable only for very large values of $M$. The intersection observed in $\sigma^{B}_{wa}$ (bottom-left panel of fig.~\ref{fig:corya2}) in the cases $M / (\pi T) = 2$ and $M / (\pi T) \to \infty$ is also observed in fig.~\ref{fig:probe_w}, and the comment mentioned in Case 2 applies also in this case. The corresponding intersection in $\sigma^B_{aa}$ (top-right panel of fig.~\ref{fig:corya2}) is not observed in fig.~\ref{fig:probe_w}, as it happens at $\mu_w / (\pi T) \simeq 0.7$ while the curves of fig.~\ref{fig:probe_w} are made for $\mu_w / (\pi T)  = 0.2$, which is a much smaller value. Finally, $\sigma^{B}_{\varepsilon a}$ shows a quadratic dependence on $\mu_w / (\pi T)$, with a higher curvature for higher values of $M$.
			\end{itemize}


			\subsection{Conductivities controlling the response to $\vec{B}_w$}
			\label{subsec:Bw}

			Similar to the previous cases, the source term of $\vec{B}_w$ is now switched on and the remaining sources are set to zero.
			
\begin{figure}[htbp]
				\centering
				\includegraphics[width=.4\textwidth]{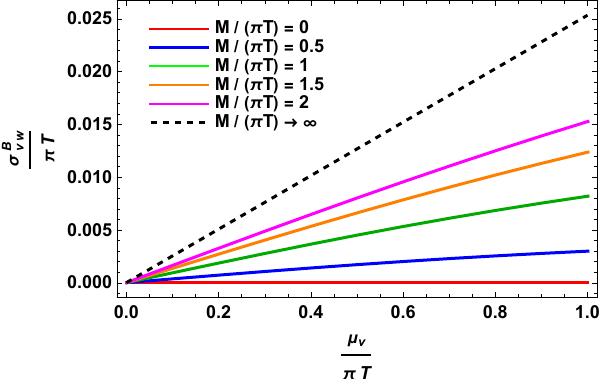}
				\qquad
				\includegraphics[width=.4\textwidth]{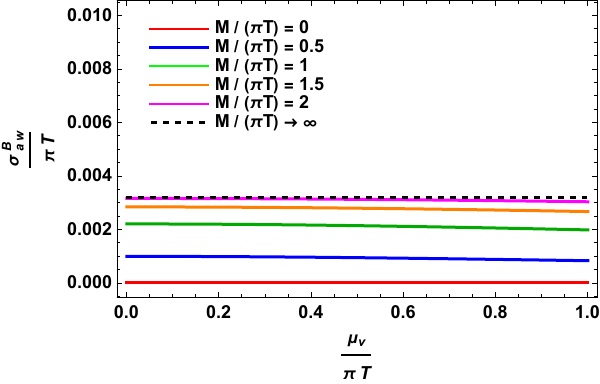}
				\qquad
				\includegraphics[width=.4\textwidth]{www}
				\qquad
				\includegraphics[width=.4\textwidth]{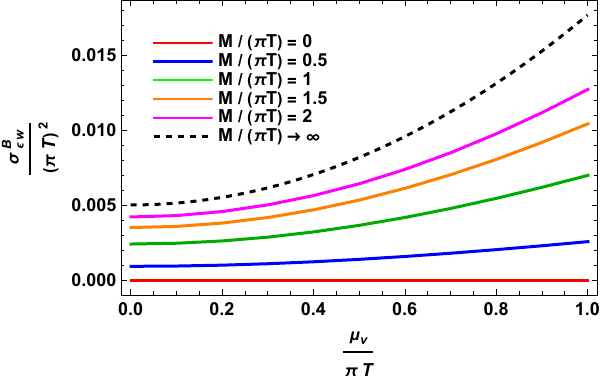}
				\caption{Case 1: Plots of the conductivities $\sigma^B_{s w}$, where $S \in \{ v, a, w, \varepsilon \}$, as a function of to $\mu_v/(\pi T)$. We have set $\mu_a/(\pi T)=0.3$ and $\mu_w/(\pi T)=0.2$.  \label{fig:coryw}}
			\end{figure}

\begin{figure}[htbp]
				\centering
				\includegraphics[width=.4\textwidth]{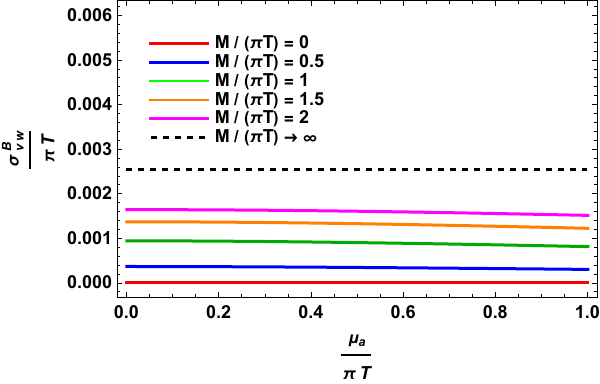}
				\qquad
				\includegraphics[width=.4\textwidth]{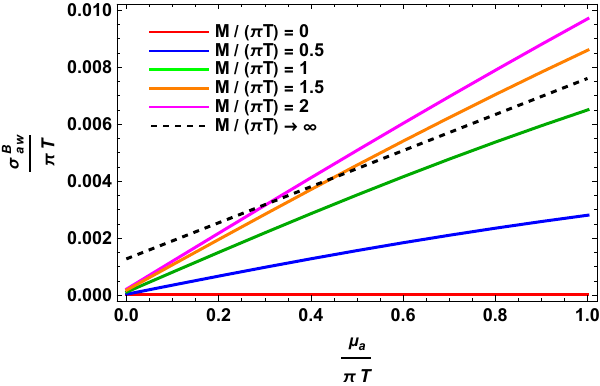}
				\qquad
				\includegraphics[width=.4\textwidth]{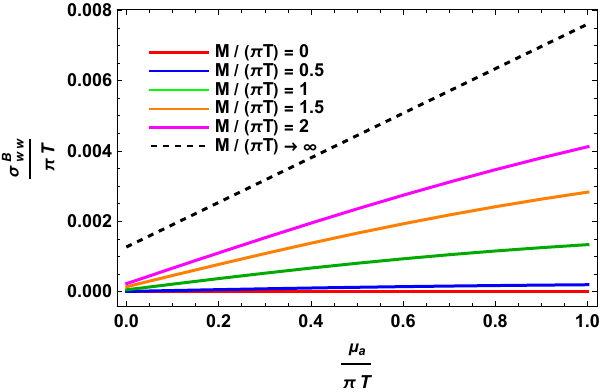}
				\qquad
				\includegraphics[width=.4\textwidth]{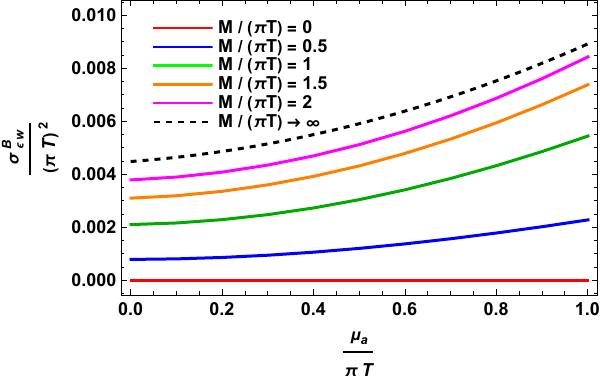}
				\caption{Case 2: Plots of the conductivities $\sigma^B_{s w}$, where $s \in \{ v, a, w, \varepsilon \}$, as a function of to $\mu_a/(\pi T)$. We have set $\mu_v/(\pi T)=0.1$ and $\mu_w/(\pi T)=0.2$. \label{fig:coryw1}}
			\end{figure}

\begin{figure}[htbp]
				\centering
				\includegraphics[width=.4\textwidth]{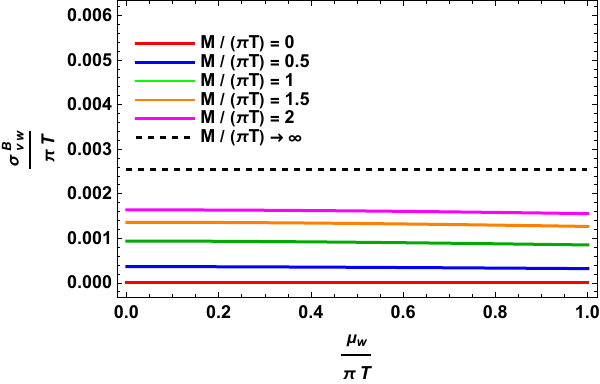}
				\qquad
				\includegraphics[width=.4\textwidth]{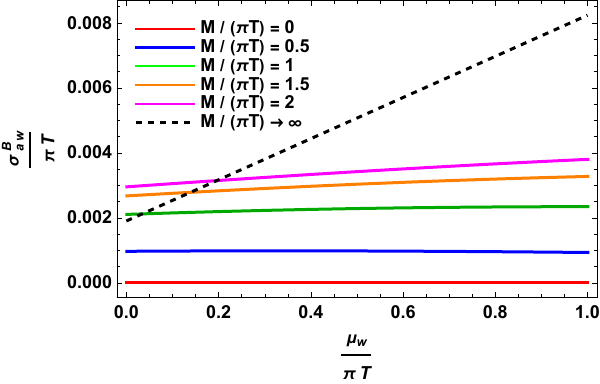}
				\qquad
				\includegraphics[width=.4\textwidth]{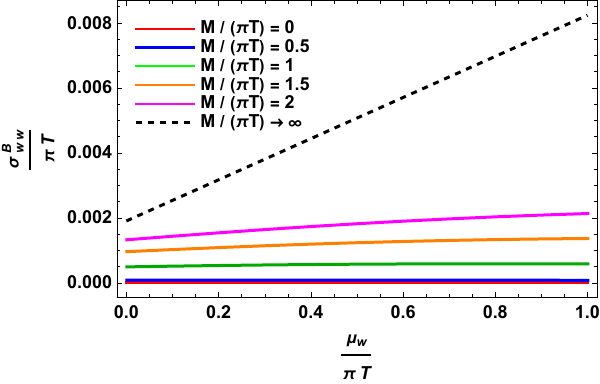}
				\qquad
				\includegraphics[width=.4\textwidth]{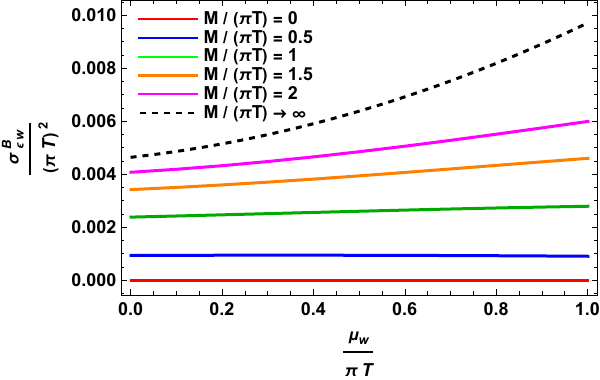}
				\caption{Case 3: Plots of the conductivities $\sigma^B_{s w}$, where $s \in \{ v, a, w, \varepsilon \}$, as a function of $\mu_w/(\pi T)$. We have set  $\mu_v/(\pi T)=0.1$ and $\mu_a/(\pi T)=0.3$. \label{fig:coryw2}}
			\end{figure}

			\begin{itemize}
				\item \textbf{Case 1:} The results are displayed in fig.\;\ref{fig:coryw}. The conductivity $\sigma^{B}_{vw}$ exhibits a linear growth with $\mu_v/(\pi T)$, with a slope that increases as $M$ increases. The conductivities $\sigma^{B}_{aw}$ and $\sigma^{B}_{ww}$, on the other hand, remain constant with respect to $\mu_v/(\pi T)$, though their magnitudes shift upward with increasing $M$. Finally, we find for $\sigma^{B}_{\varepsilon w}$ a quadratic dependence on $\mu_v/(\pi T)$, with an increasing curvature as $M$ increases.

				\item \textbf{Case 2:} The results are displayed in fig.\;\ref{fig:coryw1}. It can be seen that $\sigma^{B}_{vw}$ remains constant with $\mu_a/(\pi T)$, but its value increases steadily with $M$. The conductivities $\sigma^{B}_{aw}$ and $\sigma^{B}_{ww}$ grow linearly with $\mu_a/(\pi T)$, with their slopes enhanced as $M$ increases, except for the case $\sigma^B_{aw}$ and very large values of $M$, as it was discussed in Case 2 of sec.~\ref{subsec:Ba}. Finally, $\sigma^{B}_{\varepsilon w}$ shows a quadratic dependence on $\mu_a/(\pi T)$, with an increasing curvature as $M$ increases, similar to Case 1.  
				
				\item \textbf{Case 3:} The results are displayed in fig.~\ref{fig:coryw2}. The conductivity $\sigma^{B}_{vw}$ remains independent of $\mu_w/(\pi T)$, but it value increases steadily with $M$. The conductivities $\sigma^{B}_{aw}$ and $\sigma^{B}_{ww}$, on the other hand, grow linearly with $\mu_w/(\pi T)$, with their slopes enhanced as $M$ increases. The intersection observed for $\sigma^B_{aw}$ in the top-right panel of fig.~\ref{fig:coryw2} has the same discussion as in Case 3 of sec.~\ref{subsec:Ba}. Finally, $\sigma^B_{\varepsilon w}$ has a linear $+$ quadratic dependence on $\mu_w / (\pi T)$ with higher curvature for higher values of $M$. For very large values of $M$, this dependence is of the form $\sigma^B_{\varepsilon w} \propto (\mu_a + \mu_w)^2$, as it can be seen in eq.~(\ref{eq:App_sigmaABw_ana_M}).
			\end{itemize}

			\subsection{Conductivities controlling the  response to $\vec{\Omega}$}
			\label{subsec:Omega}

			Similar to the previous cases, the source term of the vorticity vector $\vec{\Omega}$ is now switched on and the remaining sources are set to zero.

\begin{figure}[htbp]
				\centering
				\includegraphics[width=.4\textwidth]{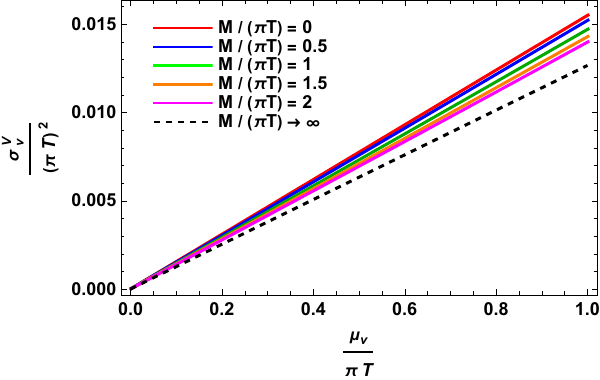}
				\qquad
				\includegraphics[width=.4\textwidth]{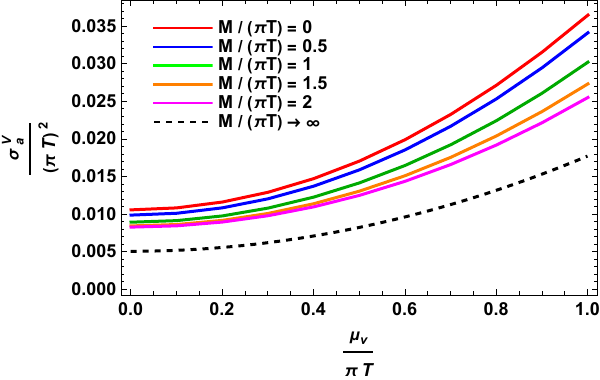}
				\qquad
				\includegraphics[width=.4\textwidth]{wee}
				\qquad
				\includegraphics[width=.4\textwidth]{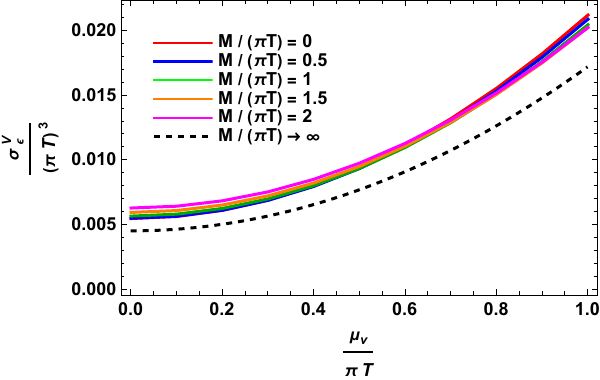}
				\caption{Case 1: Plots of the conductivities $\sigma^V_{s}$, where $s\in \{ v, a, w, \varepsilon \}$, as a function of $\mu_v/(\pi T)$. We have set $\mu_a/(\pi T)=0.3$ and $\mu_w/(\pi T)=0.2$.} 
				\label{fig:corye}
			\end{figure}

\begin{figure}[htbp]
				\centering
				\includegraphics[width=.4\textwidth]{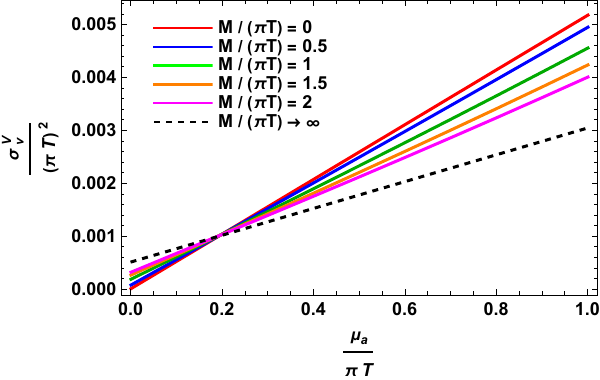}
				\qquad
				\includegraphics[width=.4\textwidth]{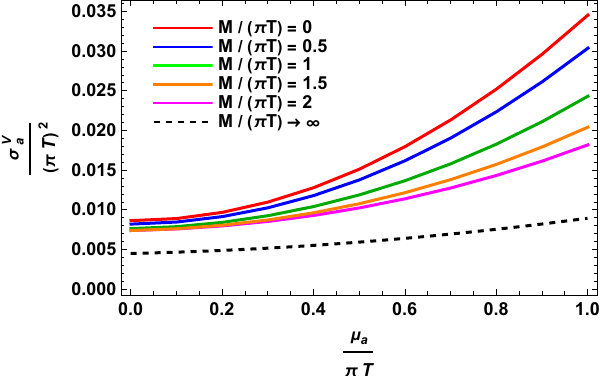}
				\qquad
				\includegraphics[width=.4\textwidth]{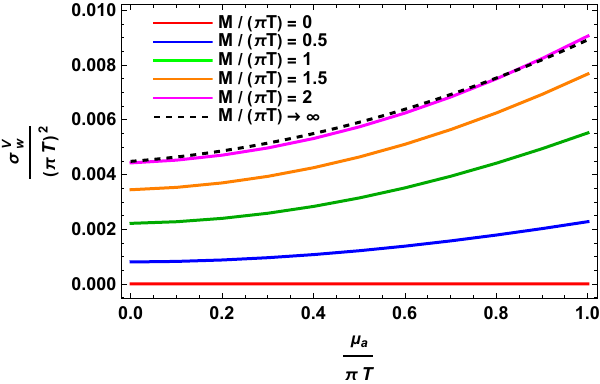}
				\qquad
				\includegraphics[width=.4\textwidth]{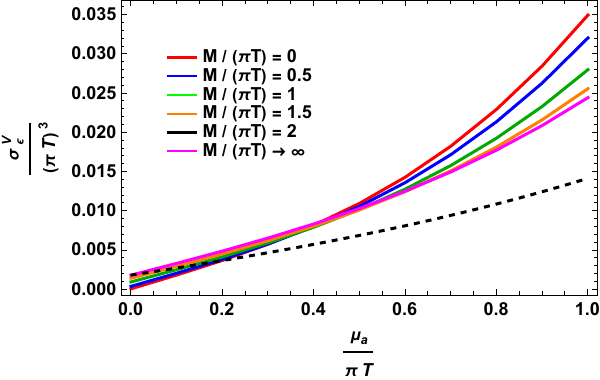}
				\caption{Case 2: Plots of the conductivities $\sigma^V_{s}$, where $s \in \{ v, a , w, \varepsilon \}$, as a function of $\mu_a/(\pi T)$. We have set $\mu_v/(\pi T)=0.1$ and $\mu_w/(\pi T)=0.2$.} 
				\label{fig:corye1}
			\end{figure}

\begin{figure}[htbp]
				\centering
				\includegraphics[width=.4\textwidth]{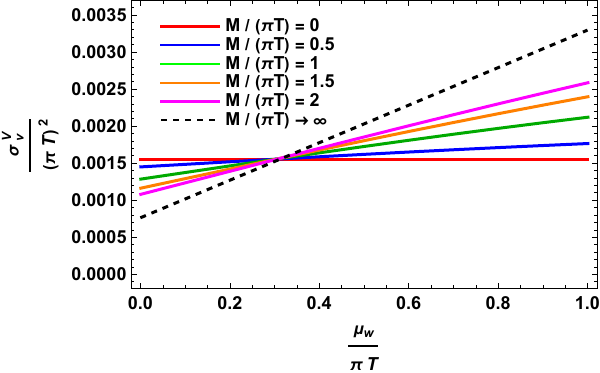}
				\qquad
				\includegraphics[width=.4\textwidth]{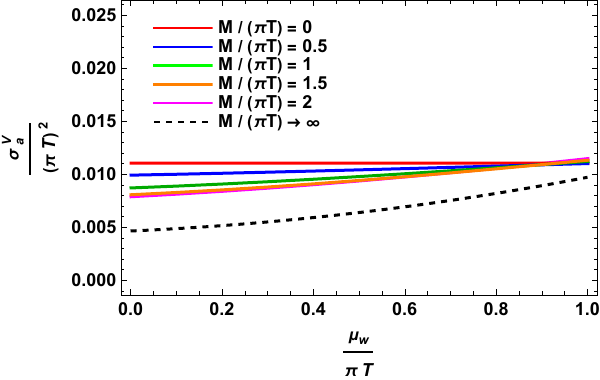}
				\qquad
				\includegraphics[width=.4\textwidth]{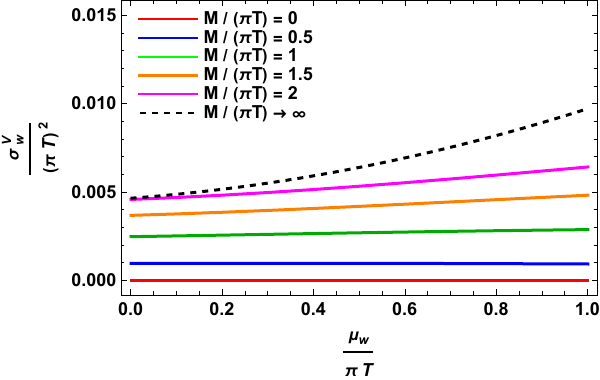}
				\qquad
				\includegraphics[width=.4\textwidth]{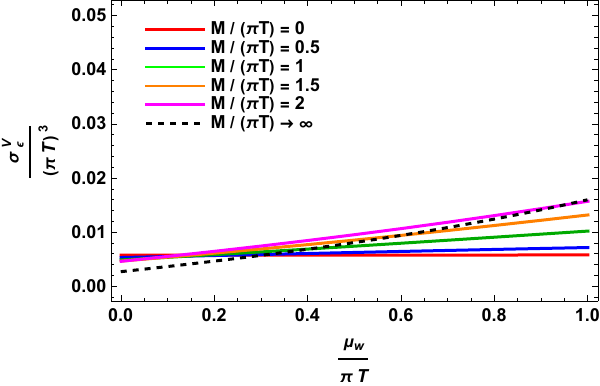}
				\caption{Case 3: Plots of the conductivities $\sigma^V_{s}$, where $s \in \{ v, a, w, \varepsilon \}$, as a function of $\mu_w/(\pi T)$. We have set $\mu_v/(\pi T)=0.1$ and $\mu_a/(\pi T)=0.3$. \label{fig:corye2}}
			\end{figure}

			\begin{itemize}
				\item \textbf{Case 1:} The results are displayed in fig.\;\ref{fig:corye}. The conductivity $\sigma^{V}_{v}$ increases linearly with $\mu_v/(\pi T)$, and its slope decreases with $M$. The conductivities $\sigma^{V}_{a}$ and $\sigma^{V}_{w}$ exhibit a quadratic dependence on $\mu_v/(\pi T)$, with their curvatures decreasing and increasing respectively as $M$ increases. Finally, $\sigma^{V}_{\varepsilon}$ shows also a quadratic dependence with $\mu_v/(\pi T)$, with slightly smaller curvature for higher values of $M$.

				\item \textbf{Case 2:} The results are displayed in fig.~\ref{fig:corye1}. The conductivity $\sigma^{V}_{v}$ shows also a linear dependence with $\mu_a/(\pi T)$, with a slope that decreases with $M$. The conductivities $\sigma^{V}_{v}$ and $\sigma^{V}_{w}$ have a quadratic dependence on $\mu_a/(\pi T)$, with their curvature decreasing and increasing, respectively, as $M$ increases. Finally, $\sigma^{V}_{\varepsilon }$ follows a linear (at small $\mu_a / (\pi T)$) and cubic (at large values of $\mu_a / (\pi T)$) dependence on $\mu_a/(\pi T)$. For very large values of $M$, this dependence is of the form $\sigma^{V}_{\varepsilon} \big|_{\mu_a \ll T} \propto \mu_v^2 (\mu_a + \mu_w)$ and $\sigma^{V}_{\varepsilon} \big|_{\mu_a \gg T} \propto (\mu_a + \mu_w)^3$, as it can be seen in eq.~(\ref{eq:App_sigmaAV_ana_M}).
				
				\item \textbf{Case 3:} The results are displayed in fig.~\ref{fig:corye2}. The conductivity $\sigma^{V}_{v}$ shows a linear dependence with $\mu_w / (\pi T)$, with a slope that increases as $M$ increases. The conductivities $\sigma^{V}_{a}$ and $\sigma^V_{w}$ exhibit a linear $+$ quadratic dependence on $\mu_w / (\pi T)$, with a curvature that increases as $M$ increases in both cases. Finally, $\sigma^V_{\varepsilon}$ shows a linear (at small $\mu_w / (\pi T)$) and cubic (at large values of $\mu_w / (\pi T)$) dependence on $\mu_w / (\pi T)$, a behavior that is confirmed by eq.~(\ref{eq:App_sigmaAV_ana_M}) in the large $M$ limit.

			\end{itemize}

\newpage

			\section{Conductivities in the large $M$ limit}
			\label{sec:App_large_M}

			We present in this appendix the explicit expressions of the conductivities in the large $M$ limit and in the basis $(V,A,W)$. These can be easily inferred from eqs.~(\ref{eq:sigma_Minfty}), (\ref{eq:sigmaAB_ana_M})-(\ref{eq:sigmaAV_ana_M}). The results~are
			\begin{eqnarray}
				\sigma^{B}_{s v}(\infty) &=& \frac{1}{4\pi^2} \left\{
				\begin{array}{ll}
					\mu_a + \mu_w \,, & \qquad (s = v) \\
					\mu_v  \,, &  \qquad (s = a) \\
					\mu_v  \,, &  \qquad (s = w) \\
					\mu_v (\mu_a + \mu_w) \,, & \qquad (s = \varepsilon)
				\end{array}
				\right.,   \label{eq:App_sigmaAB_ana_M} \\
				\sigma^{B}_{s a}(\infty) &=& \frac{1}{4\pi^2} \left\{
				\begin{array}{ll}
					\mu_v \,, & \qquad (s = v) \\
					\frac{1}{4} (\mu_a + \mu_w) \,, & \qquad (s = a) \\
					\frac{1}{4} (\mu_a + \mu_w) \,, & \qquad (s = w) \\
					\frac{1}{2} \left(  \mu_v^2 + (\mu_a + \mu_w)^2/4 \right)  + \frac{\pi^2}{6} T^2 \,, & \qquad (s = \varepsilon)
				\end{array}
				\right.,  \label{eq:App_sigmaABa_ana_M} \\
				\sigma^{B}_{s w}(\infty) &=& \frac{1}{4\pi^2} \left\{
				\begin{array}{ll}
					\mu_v \,, & \qquad (s = v) \\
					\frac{1}{4} (\mu_a + \mu_w) \,, & \qquad (s = a) \\
					\frac{1}{4} (\mu_a + \mu_w) \,, & \qquad (s = w) \\
					\frac{1}{2} \left(  \mu_v^2 + (\mu_a + \mu_w)^2/4 \right)  + \frac{\pi^2}{6} T^2 \,, & \qquad (s = \varepsilon)
				\end{array}
				\right.,  \label{eq:App_sigmaABw_ana_M} \\
				\sigma^{V}_{s}(\infty) &=& \frac{1}{4\pi^2} \left\{
				\begin{array}{ll}
					\mu_v (\mu_a + \mu_w)\,, & \qquad (s = v) \\
					\frac{1}{2} \left(  \mu_v^2 + (\mu_a + \mu_w)^2/4 \right)  + \frac{\pi^2}{6} T^2  \,, & \qquad (s = a) \\
					\frac{1}{2} \left(  \mu_v^2 + (\mu_a + \mu_w)^2/4 \right)  + \frac{\pi^2}{6} T^2  \,, & \qquad (s = w) \\
					\frac{1}{3} (\mu_a + \mu_w) \left( 3 \mu_v^2 + (\mu_a + \mu_w)^2/4 + \pi^2 T^2 \right)  \,, & \qquad (s = \varepsilon)
				\end{array}
				\right..     \label{eq:App_sigmaAV_ana_M}
			\end{eqnarray}
			These results can be compared with the results of the conductivities in the massless case, cf. eqs.~(\ref{eq:sigmaAB_ana})-(\ref{eq:sigmaAV_ana}).
			
		\end{appendices}

		\bibliographystyle{unsrt}
		\bibliography{mybibnotes}

	\end{document}